\documentclass[aps,prd,preprintnumbers,superscriptaddress,nofootinbib]{revtex4}%
\usepackage{bm,latexsym,amsmath,amssymb,amsfonts,mathrsfs}
\usepackage[pdftex]{color}
\input{colordvi.tex}
\usepackage[pdftex,hyperfootnotes=false]{hyperref}
\usepackage{hyperref}
\usepackage{url}
\hypersetup{
    colorlinks=true,
    citecolor=cyan,
}
\newcommand*{\D}{{\rm d}}

\newcommand*{\dd}{\text{\large$\cdot$}}
\begin{document}

\title{%
UV sensitive one-loop matter power spectrum in degenerate higher-order scalar-tensor theories
}

\author{Shin'ichi Hirano}
\email[Email: ]{hirano.shinichi``at''a.mbox.nagoya-u.ac.jp}
\affiliation{Division of Particle and Astrophysical Science,
Graduate School of Science, Nagoya University, Aichi 464-8602, Japan}

\author{Tsutomu Kobayashi}
\email[Email: ]{tsutomu``at''rikkyo.ac.jp}
\affiliation{Department of Physics, Rikkyo University, Toshima, Tokyo 171-8501, Japan}

\author{Daisuke Yamauchi}
\email[Email: ]{yamauchi``at''jindai.jp}
\affiliation{Faculty of Engineering, Kanagawa University, Kanagawa, 221-8686, Japan}

\author{Shuichiro Yokoyama}
\email[Email: ]{shu``at''kmi.nagoya-u.ac.jp}
\affiliation{Kobayashi Masukawa Institute, Nagoya University, Aichi 464-8602, Japan}
\affiliation{Kavli IPMU (WPI), UTIAS, The University of  Tokyo, Kashiwa, Chiba 277-8583, Japan}

\preprint{RUP-20-27}

\begin{abstract}
We study matter density perturbations
up to third order and the one-loop matter power spectrum
in degenerate higher-order scalar-tensor (DHOST) theories
beyond Horndeski.
We systematically solve gravitational field equations and
fluid equations order by order, and find
three novel shape functions characterizing the third-order solution
in DHOST theories.
A complete form of the one-loop matter power spectrum
is then obtained using the resultant second- and third-order solutions.
We confirm the previous result that the convergence condition of the loop integrals in the infrared limit becomes more stringent
than that of the standard one in general relativity.
We show that also in the ultraviolet limit the convergence condition becomes more stringent
and the one-loop matter power spectrum is thus sensitive to the short-wavelength behavior of the linear power spectrum.
\end{abstract}
\maketitle

\section{Introduction}

Scalar-tensor theories are exciting candidates of the origin of the accelerated cosmic expansion at late time~\cite{Clifton:2011jh,Heisenberg:2018vsk}.
While the scalar degree of freedom causes the accelerated expansion on large cosmological scales, 
its effect on small scale gravity experiments
must be highly suppressed in successful theories to evade existing stringent tests such as in the Solar System.
To achieve this, screening mechanisms are
implemented to elaborated scalar-tensor theories.
For example, the Vainshtein mechanism hides the scalar-mediated force
very effectively
through the non-linear derivative interaction of the scalar degree
of freedom~\cite{Vainshtein:1972sx}.
In the Horndeski family of theories~\cite{Horndeski:1974wa,Deffayet:2011gz,Kobayashi:2011nu},
which spans the most general scalar-tensor theory
having second-order field equations and hence is a class of theories free of Ostrogradsky's ghost~\cite{Ostro1,Woodard:2015zca},
the Vainshtein screening mechanism is implemented naturally as the Lagrangian contains powers of second derivatives of
the scalar field~\cite{Kimura:2011dc,Narikawa:2013pjr,Koyama:2013paa}.
Extending the Horndeski theory even further,
degenerate higher-order scalar-tensor (DHOST) theories have been
developed recently~\cite{Langlois:2015cwa,Crisostomi:2016czh,Achour:2016rkg}
(see Refs.~\cite{Langlois:2017mdk,Langlois:2018dxi,Kobayashi:2019hrl} for a review).
The field equations in such theories are apparently of higher order, but a careful counting of the degrees of freedom
shows that there in fact are one scalar and two tensor degrees of freedom due to the degeneracy of the theories, and consequently Ostrogradsky's ghost is removed.
New types of non-linear derivative interactions arise
in DHOST theories beyond Horndeski.
Their effect on the screening mechanism has been discussed in~\cite{Kobayashi:2014ida,Crisostomi:2017lbg,Langlois:2017dyl,Dima:2017pwp,Crisostomi:2017pjs,Hirano:2019scf,Crisostomi:2019yfo},
emphasizing that
partial breaking of Vainshtein screening
(first discovered in Ref.~\cite{Kobayashi:2014ida})
occurs in the presence of matter~\cite{Koyama:2015oma,Saito:2015fza,Sakstein:2015zoa,Sakstein:2015aac,Jain:2015edg,Babichev:2016jom,Sakstein:2016lyj,Sakstein:2016oel,Sakstein:2016ggl,Saltas:2018mxc,Kobayashi:2018xvr,Babichev:2018rfj,Saltas:2019ius,Rosyadi:2019hdb,Anson:2020fum,Banerjee:2020rrd}.
The non-standard interactions between
scalar and gravitational-wave degrees of freedom
in DHOST theories result also in the decay of gravitons~\cite{Creminelli:2018xsv,Creminelli:2019nok}.

In this paper, we study the impact of the non-linear derivative interactions
of the scalar field in DHOST theories on
non-linear evolution of matter density perturbations.
It has been found that the matter bispectrum in a class of DHOST theories
shows a distinct feature at the equilateral and folded limits
compared with that in general relativity (GR)~\cite{Hirano:2018uar}.
Refs.~\cite{Crisostomi:2019vhj,Lewandowski:2019txi}
have investigated the bi and tri-spectra of the matter
density perturbations
and also the one-loop matter power spectrum
in the context of DHOST theories.
They have, in particular, focused on the behaviors at the infrared (IR) limit: the squeezed limit for the bispectrum,
the double soft limit for the trispectrum,
and the IR contributions in the loop integrals
for the one-loop power spectrum,
and studied
a consistency relation for large scale structure and its violation.
So far, in DHOST theories, the third-order solution for the matter density perturbations
has been obtained only at the IR limit as done in Refs.~\cite{Crisostomi:2019vhj,Lewandowski:2019txi},
and the complete form of the one-loop matter power spectrum has not been derived yet.
The goal of this paper is therefore
to derive the third-order solution for the matter density perturbations
and to investigate the one-loop matter power spectrum in its complete form.

This paper is organized as follows.
In the next section, we introduce quadratic DHOST theories
which we focus on in this paper,
and derive basic equations for cosmological perturbations
under the quasi-static approximation.
Then, we review the solutions for the matter density perturbations
up to second order.
In Sec.~\ref{sec: third order}, we obtain the third-order solution to calculate the complete form of the one-loop matter power spectrum.
In Sec.~\ref{sec: one-loop}, we derive the one-loop matter power spectrum and investigate the asymptotic behavior of the loop integrals at the IR and ultraviolet (UV) limits.
In particular, we emphasize that the convergence condition of the loop integrals in the UV limit becomes more stringent.
Finally, we draw the conclusion of the present paper
in Sec.~\ref{sec: conclusion}.

\section{Large Scale Structure in quadratic DHOST theories}\label{sec: DHOST}

\subsection{Quadratic DHOST theories}

The action of the quadratic DHOST theories~\cite{Langlois:2015cwa} is given by
\begin{align}
    S &= \int{\rm d}^{4}x \sqrt{-g} \Bigg[ 
    G_{2}(\phi,X) 
    -G_{3}(\phi,X)\Box\phi 
    + F(\phi,X)R
    +a_{1}(\phi,X)\phi_{\mu\nu}\phi^{\mu\nu} 
        +a_{2}(\phi,X)(\Box\phi)^{2}
\notag\\
    &\hspace{2.3cm} 
        +a_{3}(\phi,X)(\Box\phi)\phi^{\mu}\phi_{\mu\nu}\phi^{\nu}
        +a_{4}(\phi,X)\phi^{\mu}\phi_{\mu\rho}\phi^{\rho\nu}\phi_{\nu}
        +a_{5}(\phi,X)(\phi^{\mu}\phi_{\mu\nu}\phi^{\nu})^{2}
        \Bigg]
        \,,\label{eq: DHOST}
\end{align}
where $\phi_{\mu}=\nabla_{\mu}\phi$, $\phi_{\nu\rho}=\nabla_{\rho}\nabla_{\nu}\phi$,
and $X=-\phi_\mu\phi^\mu/2$.
To avoid the Ostrogradsky's ghosts, we impose degeneracy conditions
among the functions $F$ and $a_i~(i=1,\cdots,5)$, and hence not all these are independent.
We focus on
the class Ia DHOST theories, as they are basically healthy and can be free from instabilities
on a cosmological background~\cite{deRham:2016wji,Langlois:2017mxy}.
The class Ia degeneracy conditions~\cite{Crisostomi:2016czh} are summarized as
\begin{align}
    & a_1+a_2=0,
\quad
    \beta_2=-6\beta_1^2,
\quad
    \beta_3=-2\beta_1\left[
2(1+\alpha_{\rm H})+\beta_1(1+\alpha_{\rm T})
\right],\label{eq: degeneracy1}
\end{align}
where
\begin{align}
    M^2 &= 2(F+2Xa_1),\quad
    M^2\alpha_{\rm T} = -4Xa_1,\quad
    M^2\alpha_{\rm H} = -4X(F_{X}+a_1),\quad
    M^2\beta_1 = 2X(F_{X}-a_2+Xa_3),\quad
\notag\\
    M^2\beta_2 &= 4X[a_1+a_2-2X(a_3+a_4)+4X^2a_5],\quad
    M^2\beta_3 = -8X(F_{X}+a_1-Xa_4)
    .\label{eq: degeneracy2}
\end{align}
Here and hereafter we use the notation $F_{X}=\partial F/\partial X$.
We thus have 3 free functions in addition to $G_2$ and $G_3$.
In the Horndeski theory~\cite{Horndeski:1974wa,Deffayet:2011gz,Kobayashi:2011nu} we have 
$\alpha_{\rm H}=\beta_{1}=\beta_{2}=\beta_{3}=0$,
while in the GLPV theory~\cite{Zumalacarregui:2013pma,Gleyzes:2014dya,Gleyzes:2014qga} we have $\beta_{1}=\beta_{2}=\beta_{3}=0$.

To keep generality, in this paper we do not impose any other constraints
among the functions. In particular, we do not take into account seriously the
constraint on scalar-tensor gravity
from the propagation speed of gravitational waves,
as the energy scale observed at LIGO is close to
the cutoff scale of the effective theory if applied to dark energy~\cite{deRham:2018red}.
One should note also that in principle the gravitational wave constraints
are irrelevant to the high-redshift universe.
For these reasons it is fair to say
that there still is a room for general DHOST theories as viable dark energy models,
and it is important to seek for independent
cosmological constraints on DHOST theories.

\subsection{Perturbation equations}

We consider a spatially flat, homogeneous and isotropic background universe,
and the cosmological perturbations in the Newtonian gauge. The perturbed metric is given by 
\begin{align}
{\rm d}s^{2} = -[1+2\Phi(t,{\bm x})]{\rm d}t^{2} + a^{2}(t)[1-2\Psi(t,{\bm x})]{\rm d}{\bm x}^{2},
\label{pertmet}
\end{align}
and the perturbed scalar field is
\begin{align}
\phi (t,{\bm x})= \phi(t) + \pi(t,{\bm x}).\label{pertscalar}
\end{align}
We introduce a dimensionless variable
$Q:=H\pi/\dot\phi$, where $H=\dot a/a$ and a dot denotes differentiation with respect to $t$.
The matter density perturbation
is defined by
\begin{align} 
\rho_{\rm  m} (t,{\bm x})= \bar\rho_{\rm m}(t)[1+\delta(t,{\bm x})].
\end{align}
We consider irrotational dust as a matter content
and therefore have only a scalar mode in the velocity field $u^i$,
which is characterized by
$\theta = \partial_{i}u^{i}/aH$.

We study the quasi-static behavior of those perturbations deep inside the horizon.
Substituting Eqs.~\eqref{pertmet} and~\eqref{pertscalar} to
the action~(\ref{eq: DHOST}) and expanding it in terms of the perturbations,
one obtains the action for the perturbations~\cite{Dima:2017pwp,Hirano:2018uar,Hirano:2019scf,Crisostomi:2019yfo,Crisostomi:2019vhj,Hiramatsu:2020fcd}.
In doing so we take the quasi-static approximation 
and neglect time derivatives compared to spatial derivatives.
Note, however, that in DHOST theories we have mixed derivatives such as 
$\partial^2\dot Q$ which cannot be simply ignored~\cite{Kobayashi:2014ida}.
We thus arrive in the end at the action for the quasi-static perturbations,
which we vary to derive
the following equations of motion (EoMs)
in Fourier space:
\begin{align}
    &(1+\alpha_{\rm T})\Psi-(1+\alpha_{\rm H})\Phi+b_2 Q +\alpha_{\rm H}\frac{\dot Q}{H}
    = 
    -\frac{\alpha_{\rm T} -4\alpha_{\rm H}}{4a^{2}H^{2}p^{2}}{\cal S}_{\gamma}[t,{\bm p};Q,Q] -\frac{\alpha_{\rm H}}{a^{2}H^{2}p^{2}}{\cal S}_{\alpha_{s}}[t,{\bm p};Q,Q],
    \label{eq: EoM Psi F}
\\
    &(1+\alpha_{\rm H})\Psi-\frac{\beta_3}{2}\Phi
+b_1Q+\frac{2\beta_1+\beta_3}{2}\frac{\dot Q}{H}+\frac{a^2}{2M^2p^2}\overline\rho_{\rm m}\delta
    = 
    \frac{d_{2} +2(2\beta_{1}+\beta_{3})}{2a^{2}H^{2}p^{2}}{\cal S}_{\gamma}[t,{\bm p};Q,Q] -\frac{2\beta_{1}+\beta_{3}}{2a^{2}H^{2}p^{2}}{\cal S}_{\alpha_{s}}[t,{\bm p};Q,Q],
    \label{eq: EoM Phi F}
\\
    &c_1\Phi +c_2\Psi + b_3Q+4\alpha_{\rm H}\frac{\dot\Psi}{H}-2(2\beta_1+\beta_3)\frac{\dot\Phi}{H}
+b_4\frac{\dot Q}{H}+2(4\beta_1+\beta_3)\frac{\ddot Q}{H^2}
\notag\\
    &= 
    \frac{d_{1}}{a^{2}H^{2}p^{2}}{\cal S}_{\gamma}[t,{\bm p};Q,Q] 
    +\frac{2\alpha_{\rm T}}{a^{2}H^{2}p^{2}}{\cal S}_{\gamma}[t,{\bm p};Q,\Psi] 
    +\frac{4d_{2}}{a^{2}H^{2}p^{2}}{\cal S}_{\gamma}[t,{\bm p};Q,\Phi] 
\notag\\
    &\quad -\frac{4\alpha_{\rm H}}{a^{2}H^{2}p^{2}}{\cal S}_{\alpha}[t,{\bm p};\Psi,Q]
    +\frac{2(2\beta_{1}+\beta_{3})}{a^{2}H^{2}p^{2}} {\cal S}_{\alpha}[t,{\bm p};\Phi,Q]
    -\frac{2(4\beta_{1}+\beta_{3})}{a^{2}H^{3}p^{2}}{\cal S}_{\alpha}[t,{\bm p};Q,\dot Q]
\notag\\
    &\quad 
    -\frac{4(4\beta_{1}+\beta_{3})}{a^{2}H^{3}p^{2}}({\cal S}_{\alpha_{s}}[t,{\bm p};Q,\dot Q] -{\cal S}_{\gamma}[t,{\bm p};Q,\dot Q])
    -\frac{b_4}{a^2H^2p^2}
    ({\cal S}_{\alpha_{s}}[t,{\bm p};Q,Q] -{\cal S}_{\gamma}[t,{\bm p};Q,Q])
\notag\\
    &\quad +\frac{2d_{2} +\alpha_{\rm T}}{a^{4}H^{4}p^{2}}{\cal T}_{\xi}[t,{\bm p};Q,Q,Q] 
    -\frac{2(4\beta_{1}+\beta_{3})}{a^{4}H^{4}p^{2}}{\cal T}_{\zeta}[t,{\bm p};Q,Q,Q],
    \label{eq: EoM Q F}
\end{align}
where 
${\bm p}$ denotes a comoving wave vector in Fourier space and $p=|{\bm p}|$.
Here, 
${\cal S}_{\Sigma}[t,{\bm p};Y,Z]~(\Sigma = \alpha,\alpha_s,\gamma)$ and ${\cal T}_{\Upsilon}[t,{\bm p};X,Y,Z]~(\Upsilon = \xi,\zeta)$ are respectively second and third-order contributions with respect to the metric and scalar field perturbations, and
are defined by
\begin{align}
    {\cal S}_{\Sigma}[t,{\bm p};Y,Z] &= 
    \frac{1}{(2\pi)^{3}}\int{\rm  d}^{3}k_{1}{\rm d}^{3}k_{2}\, \delta_{\rm D}({\bm k}_{1}+{\bm k}_{2}-{\bm p}) k_{1}^{2}k_{2}^{2}\Sigma ({\bm k}_{1},{\bm k}_{2}) Y(t,{\bm k}_{1})Z(t,{\bm k}_{2}),
    \label{eq: I}
\\
    {\cal T}_{\Upsilon}[t,{\bm p};X,Y,Z] &= 
    \frac{1}{(2\pi)^{6}}\int {\rm d}^{3}k_{1}{\rm d}^{3}k_{2}{\rm d}^{3}k_{3}\, 
    \delta_{\rm D}({\bm k}_{1}+{\bm k}_{2}+{\bm k}_{3}-{\bm p})
    k_{1}^{2}k_{2}^{2}k_{3}^{2}
    \Upsilon ({\bm k}_1,{\bm k}_2,{\bm k}_3)X(t,{\bm k}_{1})Y(t,{\bm k}_{2})Z(t,{\bm k}_{3}),
    \label{eq: J}
\end{align}
where
\begin{align}
    \alpha({\bm k}_{1},{\bm k}_{2}) &= 
    1 +\frac{{\bm k}_{1}\cdot{{\bm k}_{2}}}{k_{1}^{2}},
    \label{eq: alpha}
\\
    \alpha_{s}({\bm k}_{1},{\bm k}_{2}) &=
    \frac{1}{2} \left[ \alpha ({\bm k}_1, {\bm k}_2) + \alpha ({\bm k}_2, {\bm k}_1) \right],
    \label{eq: alphas}
\\
    \gamma({\bm k}_{1},{\bm k}_{2}) &= 
    1 -\frac{({\bm k}_{1}\cdot{\bm k}_{2})^{2}}{k_{1}^{2}k_{2}^{2}},
    \label{eq: gamma}
\\
    \xi({\bm k}_1,{\bm k}_2,{\bm k}_3) &= 
    1 -3\frac{({\bm k}_{2}\cdot{\bm k}_{3})^{2}}{k_{2}^{2}k_{3}^{2}} +2\frac{({\bm k}_{1}\cdot{\bm k}_{2})({\bm k}_{2}\cdot{\bm k}_{3})({\bm k}_{3}\cdot{\bm k}_{1})}{k_{1}^{2}k_{2}^{2}k_{3}^{2}},\label{eq:xi def}
\\
    \zeta({\bm k}_1,{\bm k}_2,{\bm k}_3) &= 
    \frac{({\bm k}_{2}\cdot{\bm k}_{3})^{2} }{k_{2}^{2}k_{3}^{2}}
    +2\frac{({\bm k}_{1}\cdot{\bm k}_{3})({\bm k}_{2}\cdot{\bm k}_{3})^{2}}{k_{1}^{2}k_{2}^{2}k_{3}^{2}}
    +\frac{{\bm k}_{2}\cdot{\bm k}_{3}}{k_2^2}
    +\frac{({\bm k}_{1}\cdot{\bm k}_{2} +{\bm k}_{3}\cdot{\bm k}_{1})({\bm k}_{2}\cdot{\bm k}_{3})}{k_{1}^{2}k_{2}^{2}},\label{eq:zeta def}
\end{align}
and
$\delta_{\rm D}({\bm k})$ is the Dirac delta function.
The explicit expressions for the time-dependent coefficients
$c_{1},c_{2},c_{3}$, $b_{1},b_{2},b_{3}$, $d_{1}$, and $d_{2}$
in terms of the functions in the DHOST action are presented in
Appendix~\ref{sec:Definition of the coefficients in the equations of motion}.
The terms involving $\alpha_{\rm H}$, $\beta_{1}$, and $\beta_{3}$ are
specific to theories more general than Horndeski.
(Here, $\beta_2$ is removed by using one of the degeneracy conditions~(\ref{eq: degeneracy1}), but $\beta_3$ is retained because using $\beta_3$ leads to simpler expressions.)
In GR,
the above set of equations corresponds to
the Poisson equation and $\Phi = \Psi$ that are
used to express the matter density perturbations
$\delta$ in terms of the
metric potentials.
In scalar-tensor theories, the Poisson equation is modified and
anisotropic stress induced by the scalar field changes the relation between
$\Phi$ and $\Psi$.
Non-linear self-interactions of the scalar field also come into play in the equations.
In particular, ${\cal S}_{\alpha}$ and ${\cal T}_{\zeta}$
newly appear in DHOST theories beyond Horndeski,
and it turns out that these terms
lead to the more stringent convergence conditions of the loop integrals in the one-loop matter power spectrum
as we will see in Sec.~\ref{sec: loop}.

We assume that the matter is minimally coupled to gravity.
Then, fluid equations are the same as the standard ones in GR,
\begin{align}
    &\frac{1}{H}\frac{\partial\delta(t,{\bm p})}{\partial t} + \theta(t,{\bm p}) = 
    -\frac{1}{(2\pi)^{3}}\int{\rm d}^{3}k_{1}{\rm d}^{3}k_{2}\, \delta_{\rm D}({\bm k}_{1}+{\bm k}_{2}-{\bm p})\alpha ({\bm k}_{1},{\bm k}_{2}) \theta(t,{\bm k}_{1})\delta(t,{\bm k}_{2}),
    \label{eq: continuous F} 
\\
    &\frac{1}{H}\frac{\partial \theta(t,{\bm p})}{\partial t} + \left(2+\frac{\dot H}{H^{2}}\right)\theta(t,{\bm p}) -\frac{p^{2}}{a^{2}H^{2}}\Phi (t,{\bm p})
\notag\\
    &\hspace{3cm}=  -\frac{1}{(2\pi)^{3}}\int {\rm d}^{3}k_{1}{\rm d}^{3}k_{2}\, \delta_{\rm D}({\bm k}_{1}+{\bm k}_{2}-{\bm p})\Bigl[\alpha_{s} ({\bm k}_{1},{\bm k}_{2}) -\gamma ({\bm k}_{1},{\bm k}_{2})\Bigr] \theta(t,{\bm k}_{1})\theta(t,{\bm k}_{2}).
    \label{eq: Eular F}
\end{align}
Although these are the same as the standard ones,
the effects of modified gravity participate through
the gravitational potential $\Phi$ which is determined by
the EoMs~(\ref{eq: EoM Psi F})--(\ref{eq: EoM Q F}).
The non-linear terms in the right hand side also
modify the higher-order solutions from the standard ones in GR
because
these are induced by the linear solution that already
contains the effects of gravity modification.

\subsection{Solving the perturbation equations}\label{sec: solutions}

A solution to the EoMs~(\ref{eq: EoM Psi F})--(\ref{eq: EoM Q F})
and the fluid equations~(\ref{eq: continuous F}) and (\ref{eq: Eular F})
can be expressed as a perturbative series,
\begin{align}
    \delta = \sum_{n=1}\delta^{(n)},\quad 
    \Phi = \sum_{n=1}\Phi^{(n)},\quad \cdots,
\end{align}
where $\delta^{(n)}$, $\Phi^{(n)}$, $\cdots [={\cal O}(\epsilon^n)]$
are the $n$-th order quantities with $\delta^{(1)}$
being ${\cal O}(\epsilon)$ a quantity.

Let us now describe the systematic procedure to obtain the $n$-th order solution.
At $n$-th order, Eqs.~(\ref{eq: EoM Psi F}) and (\ref{eq: EoM Phi F})
are schematically written as
\begin{align}
    \mathsf{M} \left(\begin{array}{c}
    \Psi^{(n)} \\ \Phi^{(n)} \\
    \end{array}\right) 
      = 
    \mathsf{N}\left(\begin{array}{c}
    Q^{(n)} \\ \dot Q^{(n)}/H \\
    \end{array}\right) 
    -
    \frac{a^2\bar\rho_{\rm m}}{2M^2p^2}
    \left(\begin{array}{c}
    0 \\ \delta^{(n)} \\
     \end{array}\right) 
    -
    \frac{a^2H^2}{p^2}\mathsf{O}^{(n)}
    \left(\begin{array}{c}
    W_{\Pi_1}^{(n)} \\ \vdots \\ W_{\Pi_m}^{(n)} 
    \end{array}\right),
      \label{eq: psiphi}
\end{align}
where
$\mathsf{M}$ and $\mathsf{N}$ are $2\times2$ matrices, $\mathsf{O}^{(n)}$ is a  $2\times m$ matrix
with $m$ being different for different $n$,
and $W^{(n)}_{\Pi_1},\cdots,W^{(n)}_{\Pi_m}$ are the $n$-th order 
functions 
with respect to the initial density field, $\delta_{\rm L}$,
which come from the 
higher-order contributions in 
Eqs.~(\ref{eq: EoM Psi F}), (\ref{eq: EoM Phi F}),
(\ref{eq: EoM Q F}),
(\ref{eq: continuous F}), and (\ref{eq: Eular F}).
Here, $\Pi$ represents the shape 
of 
the kernel of the non-linear mode-coupling.
We define $W^{(n)}_\Pi$ as
\begin{align}
      W^{(n)}_\Pi ({\bm p}) &:= 
      \int\frac{{\rm d}^3k_1\cdots{\rm d}^3k_n}{(2\pi )^{3(n-1)}} \delta_{\rm D}\left({\bm k}_1+\cdots +{\bm k}_n-{\bm p}\right)
      \Pi({\bm k}_1,\cdots,{\bm k}_n)\delta_{\rm L}({\bm k}_1)\cdots\delta_{\rm L}({\bm k}_n)
      ,\label{eq:W_Pi def}
\end{align}
where $\delta_{\rm L}({\bm k})$ denotes the initial density field.
The matrices $\mathsf{M}$ and $\mathsf{N}$ are independent of $n$,
whose explicit forms are presented in Appendix~\ref{sec: coeff homogeneous}.
 Inverting $\mathsf{M}$, we obtain the solutions for $\Psi^{(n)}$ and $\Phi^{(n)}$ as follows:
 \begin{align}
    \left(\begin{array}{c}
    \Psi^{(n)} \\ \Phi^{(n)} \\
         \end{array}\right) 
      =
    \mathsf{M}^{-1} \mathsf{N}
    \left(\begin{array}{c}
    Q^{(n)} \\ \dot Q^{(n)}/H \\
    \end{array} \right) 
    -
    \frac{a^2\bar\rho_{\rm m}}{2M^2p^2}\mathsf{M}^{-1}
    \left(\begin{array}{c}
    0 \\ \delta^{(n)} \\
    \end{array}\right) 
    -
    \frac{a^2H^2}{p^2}\mathsf{M}^{-1}\mathsf{O}^{(n)}
    \left(\begin{array}{c}
    W_{\Pi_1}^{(n)} \\ \vdots \\ W_{\Pi_m}^{(n)} \\
    \end{array}\right).
    \label{eq: sol psiphi}
\end{align}
 Substituting these and their time derivatives into
 the $n$-th order part of
 Eq.~(\ref{eq: EoM Q F}), we obtain the solution for $Q^{(n)}$:
\begin{align}
     Q^{(n)} &= -\frac{a^{2}H^{2}}{p^{2}}\Bigg(\nu_{Q} \frac{\dot{\delta}^{(n)}}{H} + \kappa_{Q}\delta^{(n)}\Bigg)
    -\frac{a^{2}H^{2}}{p^{2}}\sum_{\Pi\in U_n}\tau^{(n)}_{Q,\Pi}W^{(n)}_\Pi,
    \label{eq: Q F_{n}}
\end{align}
where $U_n=\{\Pi_1\cdots ,\Pi_m\}$ denotes the set of the kernels to describe the $n$-th order mode-coupling.
At this step $\dot Q^{(n)}$ and $\ddot Q^{(n)}$
are all cancelled thanks to the degeneracy conditions,
so that the equation can be solved algebraically for $Q^{(n)}$.
The explicit forms of the coefficients
$\nu_{Q}$ and $\kappa_{Q}$ are given in Appendix~\ref{sec: coeff homogeneous}.
It should be noted that
$\dot \delta^{(n)}$ appears in Eq.~\eqref{eq: Q F_{n}}
when one considers theories more general than Horndeski~\cite{Kobayashi:2014ida,DAmico:2016ntq,Hirano:2018uar,Hirano:2019scf,Crisostomi:2019yfo,Crisostomi:2019vhj,Hiramatsu:2020fcd}.
Given the concrete form of $\mathsf{O}^{(n)}$,
it is straightforward to write the explicit form of $\tau^{(n)}_{Q,\Pi}$.
Turning back to Eq.~(\ref{eq: sol psiphi}),
now one can eliminate $Q^{(n)}$ and $\dot Q^{(n)}$ in the right hand side
to express
$\Psi^{(n)}$ and $\Phi^{(n)}$ in terms of $\delta^{(n)}$
and $W_\Pi^{(n)}$ as 
 \begin{align}
     \Psi^{(n)} &= 
    -\frac{a^{2}H^{2}}{p^{2}}\Bigg(\mu_{\Psi} \frac{\ddot{\delta}^{(n)}}{H^{2}} +\nu_{\Psi} \frac{\dot{\delta}^{(n)}}{H} + \kappa_{\Psi}\delta^{(n)}\Bigg)
    -\frac{a^{2}H^{2}}{p^{2}}\sum_{\Pi\in U_n}\tau^{(n)}_{\Psi ,\Pi}W^{(n)}_\Pi,
    \label{eq: Psi F_{n}}
\\
    \Phi^{(n)} &= 
    -\frac{a^{2}H^{2}}{p^{2}}\Bigg(\mu_{\Phi} \frac{\ddot{\delta}^{(n)}}{H^{2}}  +\nu_{\Phi} \frac{\dot{\delta}^{(n)}}{H} + \kappa_{\Phi}\delta^{(n)}\Bigg)
    -\frac{a^{2}H^{2}}{p^{2}}\sum_{\Pi\in U_n}\tau^{(n)}_{\Phi ,\Pi}W^{(n)}_\Pi.
    \label{eq: Phi F_{n}}
 \end{align}
The explicit expressions for the coefficients of the homogeneous solutions,
$\mu_\Psi, \nu_\Psi,\cdots$, are also
given in Appendix~\ref{sec: coeff homogeneous}. We also show the coefficients of the second- and third-order mode-couplings in Appendices \ref{sec:Coefficients of higher-order solutions}--\ref{sec: coeff third}.

Having thus obtained $\Psi^{(n)}$, $\Phi^{(n)}$, and $Q^{(n)}$
expressed in terms of the matter density perturbations,
we use the fluid equations~(\ref{eq: continuous F}),~(\ref{eq: Eular F}),
 and~(\ref{eq: Phi F_{n}}) to obtain the evolution equation for $\delta^{(n)}$:
 \begin{align}
    &\frac{\partial^{2} \delta^{(n)}}{\partial t^{2}} + (2+\varsigma)H\frac{\partial \delta^{(n)}}{\partial t} -\frac{3}{2}\Omega_{\mathrm{m}}\Xi_{\Phi}H^{2}\delta^{(n)}= H^{2}\sum_{\Pi\in U_n} S^{(n)}_{\Pi}W^{(n)}_{\Pi},
    \label{eq: evolution equation}
\end{align}
where
\begin{align}
    \varsigma := \frac{2\mu_{\Phi} -\nu_{\Phi}}{1-\mu_{\Phi}},\quad 
    \frac{3}{2}\Omega_{\rm m}\Xi_{\Phi} := \frac{\kappa_{\Phi}}{1-\mu_{\Phi}},\quad
    \Omega_{\rm m} := \frac{\bar\rho_{\rm m}}{3M^2H^2}.
    \label{eq:Omega_m def}
\end{align}
We assume that there is no intrinsic non-linearity at an initial time $t_{\rm i}$,
{\it i.e.},
$\delta^{(2)}(t_{\rm i})=\delta^{(3)}(t_{\rm i}) = \cdots =0$.
Then, the first-order solution is obtained by
solving the homogeneous equation,
and the higher-order solutions are given solely by the inhomogeneous solution.

The first- and second-order solutions for the matter density perturbations
in DHOST theories
have already been obtained in the literature~\cite{,Crisostomi:2019vhj,Lewandowski:2019txi},
but here for completeness we replicate the previous discussion.
The third-order solution, which is obtained for the first time in this paper,
is presented in the next section.

\subsubsection{First-order solution}\label{sec: linear}

From Eq.~(\ref{eq: evolution equation}), 
by setting $W_\Pi^{(n)} = 0$,
we see that
the first-order evolution equation for the density perturbation is given by
\begin{align}
    &\frac{\partial^{2} \delta^{(1)}}{\partial t^{2}} + (2+\varsigma)H\frac{\partial \delta^{(1)}}{\partial t} -\frac{3}{2}\Omega_{\mathrm{m}}\Xi_{\Phi}H^{2}\delta^{(1)} = 0.
    \label{eq: first-order eq.}
\end{align}
This equation has growing and decaying solutions, denoted, respectively,
as $D_+(t)$ and $D_-(t)$.
Discarding the decaying solution,
we write the solution as
\begin{align}
    \delta^{(1)}(t,{\bm p}) = D_{+}(t)\delta_{\rm L}({\bm p})
    \label{eq: delta_{1}}.
\end{align}
The linear growth rate, $f$, is convenient for characterizing
the growth of the matter density perturbations
and is defined by
\begin{align}
    f= \frac{\D \ln{D_{+}}}{\D \ln{a}}.
    \label{eq:f def}
\end{align}
Substituting the solution~(\ref{eq: delta_{1}})
into the continuity equation~(\ref{eq: continuous F}),
we obtain 
\begin{align}
    \theta^{(1)}(t,{\bm p}) = -f(t) \delta^{(1)}(t,{\bm p}).
    \label{eq: theta_{1}}
\end{align}

\subsubsection{Second-order solution}

In order to obtain the second-order solution of the matter density perturbations,
we need to substitute the first-order solutions
\eqref{eq: delta_{1}} and \eqref{eq: theta_{1}} into the right-hand-side of Eqs.~\eqref{eq: continuous F} and \eqref{eq: Eular F}.
The relevant kernel to describe the second-order mode-coupling functions
are $\alpha$ and $\gamma$ defined in Eqs.~\eqref{eq: alpha} and \eqref{eq: gamma},
namely $U_2=\{\alpha,\gamma\}$. 
With these non-linear mode-coupling functions \eqref{eq:W_Pi def},
the second-order evolution equation is given by
\begin{align}
    &\frac{\partial^{2} \delta^{(2)}}{\partial t^{2}} + (2+\varsigma)H\frac{\partial \delta^{(2)}}{\partial t} -\frac{3}{2}\Omega_{\mathrm{m}}\Xi_{\Phi}H^{2}\delta^{(2)} 
    =H^2\sum_{\Pi =\alpha ,\gamma}S^{(2)}_\Pi W_\Pi^{(2)}
    \label{second-order eq.}.
\end{align}
The coefficients of the second-order mode coupling are given by
\begin{align}
    (1-\mu_{\Phi})S^{(2)}_\alpha &= \tau^{(2)}_{\Phi ,\alpha} +\frac{1}{a^{2}H^{2}D_{+}}(a^{2}HD_{+}^{3}f)^{\dd},
\\
    (1-\mu_{\Phi})S^{(2)}_\gamma &= \tau^{(2)}_{\Phi ,\gamma} -D_{+}^{2}f^{2},
\end{align}
where the explicit form of $\tau^{(2)}_{\Phi,\Pi}$ is given in Appendices \ref{sec:Coefficients of higher-order solutions} and \ref{sec: coeff second}.

The second-order solution is obtained as 
\begin{align}
    \delta^{(2)}(t,{\bm p}) = D_{+}^{2}(t) \left[\kappa(t){W}^{(2)}_{\alpha}({\bm p})-\frac{2}{7}\lambda(t){ W}^{(2)}_{\gamma}({\bm p}) \right],
    \label{eq: delta_{2}}
\end{align}
where
\begin{align}
    \kappa(t) &= \frac{1}{D_{+}^{2}(t)} L\left[H^{2}S^{(2)}_\alpha\right], \quad 
    \lambda(t) = -\frac{7}{2D_{+}^{2}(t)}L\left[H^{2}S^{(2)}_\gamma\right],
    \label{eq: kappa-lambda}
\end{align}
and we defined the functional $L$ acting on a function $s$ of time as
\begin{align}
    L[s] := \int^{t}_{0} dT \frac{D_{+}(T)D_{-}(t)-D_{+}(t)D_{-}(T)}{D_{+}(T)\dot D_{-}(T) -\dot D_{+}(T)D_{-}(T)}s(T).
    \label{eq: L}
\end{align}

In the Einstein-de Sitter universe in
GR, we have
$\mu_{\Phi}=\varsigma=\tau^{(2)}_{\Phi,\alpha}=\tau^{(2)}_{\Phi,\gamma}=0$ and $\Xi_{\Phi}=1$.
We then see that $\lambda = 1$ and $\kappa=1$.
In the Horndeski theory, $\lambda$ can deviate much from the standard value,
$\lambda\neq1$, but $\kappa$ still takes the standard value, $\kappa=1$~\cite{Takushima:2013foa,Yamauchi:2017ibz}.
In DHOST theories beyond Horndeski, not only $\lambda$
but also $\kappa$ can deviate from $1$~\cite{Hirano:2018uar,Crisostomi:2019vhj,Lewandowski:2019txi}.
For $\kappa$ away from $1$,
the matter bispectrum is altered at the folded configuration in momentum space~\cite{Hirano:2018uar,Crisostomi:2019vhj}.

Substituting $\delta^{(1)}$, $\delta^{(2)}$, and $\theta^{(1)}$ into
Eq.~(\ref{eq: continuous F}), we can also obtain the second-order velocity divergence,
\begin{align}
    \theta^{(2)}(t,{\bm p}) &= -D_{+}^{2}f \left[\kappa_{\theta}(t){W}^{(2)}_{\alpha}({\bm p}) -\frac{4}{7}\lambda_{\theta}(t){W}^{(2)}_{\gamma}({\bm p}) \right],
    \label{eq: theta_{2}}
\end{align}
where we defined
\begin{align}
    \kappa_{\theta}(t) &= 2\kappa - 1 + \frac{\dot{\kappa}}{fH}, \\
    \lambda_{\theta}(t) &= \lambda +\frac{\dot{\lambda}}{2fH}.
\end{align}
In the Einstein de Sitter universe in GR,
$\lambda_{\theta} = 1$ and $\kappa_{\theta}=1$.
As $\lambda$ and $\kappa$ can,
in DHOST theories
both $\lambda_\theta$ and $\kappa_\theta$
can also deviate much from $1$ and they are sensitive to the 
time derivative of $\lambda$ and $\kappa$, respectively.

\section{Third-order solution}\label{sec: third order}

Having reviewed the first- and second-order solutions,
now let us proceed to derive
the third-order solution for the matter density perturbations
in DHOST theories.
The procedure is basically the same as in the case of the second-order solution.
Before going to the detailed analysis, we need to discuss the kernels of the non-linear mode-couplings of the third-order solution.
There are several choices of the shape functions to describe the third-order solutions for $\Psi$,
$\Phi$, and $Q$.
Since the mode-couplings in Eqs.~\eqref{eq: EoM Psi F} and \eqref{eq: EoM Phi F} are determined by $\alpha_s$
and $\gamma$, the relevant kernels 
can be straightforwardly chosen to be $\alpha\alpha$, $\alpha\gamma$, $\gamma\alpha$ and $\gamma\gamma$,
which are defined in Eqs.~\eqref{eq:alphaalpha def}--\eqref{eq:gammagamma def}.
Moreover, in order to include the effect of the antisymmetric part of the kernel $\alpha$ appearing in Eqs.~\eqref{eq: EoM Q F} and \eqref{eq: continuous F}, 
the additional two kernels $\alpha\alpha_\ominus$ and $\alpha\gamma_\ominus$ defined in Eqs.~\eqref{eq:alphaalpha- def} and \eqref{eq:alphagamma- def} are needed.
In addition to these six kernels, we further consider the kernels $\xi$ and $\zeta$ defined in Eqs.~\eqref{eq:xi def} and \eqref{eq:zeta def} 
to take into account the mode-couplings from the three-point self-interaction 
terms of the scalar field perturbations in Eq.~\eqref{eq: EoM Q F}.
In summary, the set of the relevant third-order shape kernels is given by 
$U_3=\{\alpha\alpha,\alpha\gamma,\gamma\alpha,\gamma\gamma,\alpha\alpha_\ominus,\alpha\gamma_\ominus,\xi,\zeta\}$.

It then follows that
the third-order evolution equation is given by
\begin{align}
    &\ddot\delta^{(3)}+ (2+\varsigma)H\dot\delta^{(3)} -\frac{3}{2}\Omega_{\mathrm{m}}\Xi_{\Phi}H^{2}\delta^{(3)} =
    H^{2}\sum_{\Pi\in U_3}S^{(3)}_{\Pi}W^{(3)}_\Pi
    \label{eq: third-order eq.}.
\end{align} 
The coefficients of the third-order mode-couplings are
\begin{align}
	(1-\mu_{\Phi})S^{(3)}_{\alpha\alpha} 
		&= \tau^{(3)}_{\Phi ,\alpha\alpha}
    			+2D_+^3f^2\kappa_{\theta} +\frac{1}{a^2H^2}\Big[a^2H D_{+}^3f(\kappa+ \kappa_{\theta})\Bigr]^{\dd}
	,\\
	(1-\mu_{\Phi})S^{(3)}_{\alpha\gamma} 
		&=\tau^{(3)}_{\Phi ,\alpha\gamma}
			-\frac{8}{7}D_+^3f^2\lambda_{\theta}-\frac{2}{7a^2H^2}\Bigl[a^2HD_+^3f(\lambda +2\lambda_{\theta})\Bigr]^{\dd}
	,\\
	(1-\mu_{\Phi})S^{(3)}_{\gamma\alpha}
		&=\tau^{(3)}_{\Phi ,\gamma\alpha}-2D_+^3f^2\kappa_{\theta}
	,\\
	(1-\mu_{\Phi})S^{(3)}_{\gamma\gamma}
		&=\tau^{(3)}_{\Phi ,\gamma\gamma}+\frac{8}{7}D_+^3f^2\lambda_{\theta}
	,\\
    (1-\mu_{\Phi})S^{(3)}_{\alpha\alpha\ominus}
		&=\tau^{(3)}_{\Phi ,\alpha\alpha\ominus}
			+\frac{1}{a^2H^2}\Bigl[a^2HD_+^3f(\kappa-\kappa_{\theta})\Bigr]^{\dd}
	,\\
	(1-\mu_{\Phi})S^{(3)}_{\alpha\gamma\ominus}
		&=\tau^{(3)}_{\Phi ,\alpha\gamma\ominus}
				-\frac{2}{7a^2H^2}\left[a^2HD_+^3f(\lambda-2\lambda_{\theta})\right]^{\dd}
	,\\
	(1-\mu_{\Phi})S^{(3)}_\xi
		&=\tau^{(3)}_{\Phi ,\xi}
	,\\
    (1-\mu_{\Phi})S^{(3)}_\zeta
		&= \tau^{(3)}_{\Phi ,\zeta}
	,
\end{align}
where the explicit expression of $\tau^{(3)}_{\Phi,\Pi}$ is shown in Appendices \ref{sec:Coefficients of higher-order solutions} and \ref{sec: coeff third}.
Using the following relation
\begin{align}
	W_{\gamma\alpha}=\frac{1}{2}\left(W_{\alpha\gamma}+W_{\alpha\gamma\ominus}\right) +W_{\gamma\gamma}-\frac{1}{2}W_\xi,
\end{align}
one can remove $W_{\gamma\alpha}$ and absorb its coefficients into $W_{\alpha\gamma}$, $W_{\alpha\gamma\ominus}$, $W_{\gamma\gamma}$, and $W_\xi$.

Following the same step as the second-order solution, we thus arrive at the
third-order solution,
\begin{align}
    \delta^{(3)}  &= D_{+}^{3}\Bigg[d_{\alpha\alpha}W^{(3)}_{\alpha\alpha} -\frac{4}{7}d_{\alpha\gamma }W^{(3)}_{\alpha\gamma} -\frac{2}{21}d_{\gamma\gamma}W^{(3)}_{\gamma\gamma} +\frac{1}{9}d_{\xi}W^{(3)}_{\xi} +d_{\alpha\alpha\ominus}W^{(3)}_{\alpha\alpha\ominus} +d_{\alpha\gamma\ominus}W^{(3)}_{\alpha\gamma\ominus} +d_{\zeta}W^{(3)}_{\zeta}\Bigg],
    \label{eq: delta_{3}}
\end{align}
where
\begin{align}
	&d_{\alpha\alpha} = \frac{1}{D_{+}^{3}}L\left[H^{2}S^{(3)}_{\alpha\alpha}\right]
	,\quad 
	d_{\alpha\gamma} = -\frac{7}{4D_{+}^{3}}L\left[H^2\left(S^{(3)}_{\alpha\gamma}+\frac{1}{2}S_{\gamma\alpha}^{(3)}\right)\right]
	,\notag\\
	&d_{\gamma\gamma} = -\frac{21}{2D_{+}^{3}}L\left[H^{2}\left(S^{(3)}_{\gamma\gamma}+S_{\gamma\alpha}^{(3)}\right)\right]
	,\quad 
	d_{\xi} = \frac{9}{D_{+}^{3}}L\left[H^{2}\left(S^{(3)}_{\xi}-\frac{1}{2}S_{\gamma\alpha}^{(3)}\right)\right]
	,\\
	&d_{\alpha\alpha\ominus} = \frac{1}{D_{+}^{3}}L\left[H^{2}S^{(3)}_{\alpha\alpha\ominus}\right]
	,\quad 
	d_{\alpha\gamma\ominus} = \frac{1}{D_{+}^{3}}L\left[H^{2}\left(S^{(3)}_{\alpha\gamma\ominus}+\frac{1}{2}S_{\gamma\alpha}^{(3)}\right)\right]
	,\quad
	d_{\zeta} = \frac{1}{D_{+}^{3}}L\left[H^{2}S^{(3)}_{\zeta}\right]
	,\notag
\end{align}
and
$L[\cdots]$ has already been defined in Eq.~(\ref{eq: L}).
In the limit of the Einstein-de Sitter universe in GR, 
it is easy to show that
$d_{\alpha\alpha}$, $d_{\alpha\gamma}$, $d_{\gamma\gamma}$, and $d_\xi$ reduce to unity,
while the other three, $d_{\alpha\alpha\ominus}$, $d_{\alpha\gamma\ominus}$, and $d_\zeta$, vanish.
In the case where gravity is described by
the Horndeski family, we still have $d_{\alpha\alpha}=1$ and $d_{\alpha\alpha\ominus}=d_{\alpha\gamma\ominus}=d_\zeta=0$,
but now $d_{\alpha\gamma}$, $d_{\gamma\gamma}$, and $d_\xi$ deviate from
unity~\cite{Takushima:2015iha,Cusin:2017wjg}.
The present analysis shows that
in DHOST theories all of these seven quantities can
have non-standard values in general.
In particular, $d_{\alpha\alpha}\neq 1$,
$d_{\alpha\alpha\ominus}\neq 0$,
$d_{\alpha\gamma\ominus}\neq 0$, and $d_\zeta\neq 0$
are specific to theories beyond Horndeski.

\section{One-loop power spectrum}\label{sec: one-loop}

We now calculate the one-loop power spectrum for 
the matter density perturbations.
The one-loop matter power spectrum has been discussed in the context of
modified gravity theories in~\cite{Koyama:2009me,Brax:2013fna,Taruya:2016jdt,Bose:2016qun},
and in particular in the context of the
Horndeski theory in~\cite{Takushima:2015iha} and DHOST
theories in~\cite{Crisostomi:2019vhj,Lewandowski:2019txi}.
In Refs.~\cite{Crisostomi:2019vhj,Lewandowski:2019txi}, the one-loop matter power spectrum in the IR limit of the loop integrals has been investigated
and
the third-order solution has been obtained only in the IR limit.
However, the complete form of the one-loop matter power spectrum 
including the UV contribution
of the loop integrals
has not been derived yet in the context of DHOST theories.
In the present paper, we calculate the complete form of the one-loop matter power spectrum by using the third-order solution derived in Sec.~\ref{sec: third order}.

The power spectrum for $\delta$ is
given in terms of the two-point correlation function as
\begin{align}
    & \langle \delta(t,{\bm k}_{1})\delta(t,{\bm k}_{2}) \rangle 
        = (2\pi)^{3}\delta({\bm k}_{1}+{\bm k}_{2})P_{\delta\delta}(t,{\bm k}_{1}).
\end{align}
In this paper, we focus on the auto-power spectrum
for $\delta$;
the one-loop cross-power spectrum between matter density perturbation
and velocity divergence, and the auto-power spectrum for velocity divergence
have the same structure as that of the auto-power spectrum for the matter density perturbation.

\subsection{One-loop matter power spectrum}

Using the solution of the matter density perturbations up to third order,
$\delta (t,{\bm k}) = \delta^{(1)}(t,{\bm k}) +\delta^{(2)}(t,{\bm k}) +\delta^{(3)}(t,{\bm k})$,
and assuming that the initial density field obeys
the Gaussian statistics, one can write
the one-loop matter power spectrum as
\begin{align}
    P_{\delta\delta}(t,{\bm k})
        &= D^{2}_{+}(t)P_{\rm L}(k) +D_{+}^{4}(t)\left[P^{(22)}_{\delta\delta}(t,{\bm k}) +2P^{(13)}_{\delta\delta}(t,{\bm k})\right],
\end{align}
where $P_{\delta\delta}^{(22)}$ and $P_{\delta\delta}^{(13)}$ are
one-loop corrections to the linear power spectrum
due to the second- and third-order solutions
defined by
\begin{align}
    & \langle \delta^{(2)}(t,{\bm k}_{1})\delta^{(2)}(t,{\bm k}_{2}) \rangle 
        = (2\pi)^{3}\delta^{(3)}({\bm k}_{1}+{\bm k}_{2})D_+^4(t)P^{(22)}_{\delta\delta}(t,{\bm k}_{1}), 
        \label{eq: P22}
\\
    & \langle \delta^{(1)}(t,{\bm k}_{1})\delta^{(3)}(t,{\bm k}_{2}) \rangle 
        = (2\pi)^{3}\delta^{(3)}({\bm k}_{1}+{\bm k}_{2})D_+^4(t)P^{(13)}_{\delta\delta}(t,{\bm k}_{1}),
        \label{eq: P13}
\end{align}
and $P_{\rm L}(k)$ is the linear power spectrum
for the initial density field $\delta_{\rm L}$.
It follows
from Eqs.~(\ref{eq: delta_{2}}) and~(\ref{eq: delta_{3}})
that the second- and third-order solutions can be written
in terms of the kernels as
\begin{align}
    \delta^{(2)}(t,{\bm k}) &= \frac{D_{+}^{2}(t)}{(2\pi)^{3}}\int {\rm d}^{3}p_{1}{\rm d}^{3}p_{2}\delta_{\rm D}({\bm p}_{1}+{\bm p}_{2}-{\bm k})
    F_{2}(t,{\bm p}_{1},{\bm p}_{2})\delta_{\rm L}({\bm p}_{1})\delta_{\rm L}({\bm p}_{2}),
\\
    \delta^{(3)}(t,{\bm k}) &= \frac{D_{+}^{3}(t)}{(2\pi)^{6}}\int {\rm d}^{3}p_{1}{\rm d}^{3}p_{2}{\rm d}^{3}p_{3}\delta_{\rm D}({\bm p}_{1}+{\bm p}_{2}+{\bm p}_{3}-{\bm k})
    F_{3}(t,{\bm p}_{1},{\bm p}_{2},{\bm p}_{3})\delta_{\rm L}({\bm p}_{1})\delta_{\rm L}({\bm p}_{2})\delta_{\rm L}({\bm p}_{3}),
\end{align}
with
\begin{align}
    F_{2}(t,{\bm p}_{1},{\bm p}_{2}) &= \kappa(t)\,  \alpha_s({\bm p}_{1},{\bm p}_{2}) -\frac{2}{7}\lambda(t)\,  \gamma({\bm p}_{1},{\bm p}_{2}),
    \label{eq: F2}
\\
    F_{3}(t,{\bm p}_{1},{\bm p}_{2},{\bm p}_{3}) &= d_{\alpha\alpha}(t)\, \alpha\alpha ({\bm p}_{1},{\bm p}_{2},{\bm p}_{3}) 
    -\frac{4}{7}d_{\alpha\gamma }(t)\, \alpha\gamma ({\bm p}_{1},{\bm p}_{2},{\bm p}_{3})
    -\frac{2}{21}d_{\gamma\gamma}(t)\, \gamma\gamma({\bm p}_{1},{\bm p}_{2},{\bm p}_{3})
    +\frac{1}{9}d_{\xi}(t)\, \xi_c({\bm p}_{1},{\bm p}_{2},{\bm p}_{3})
\notag\\
    &\quad  +d_{\alpha\alpha\ominus}(t)\, \alpha\alpha_{\ominus}({\bm p}_{1},{\bm p}_{2},{\bm p}_{3}) 
    +d_{\alpha\gamma\ominus}(t)\, \alpha\gamma_{\ominus}({\bm p}_{1},{\bm p}_{2},{\bm p}_{3})
    +d_{\zeta}(t)\, \zeta_c({\bm p}_{1},{\bm p}_{2},{\bm p}_{3}),
    \label{eq: F3}
\end{align}
where the explicit forms of the mode-coupling kernels
are shown in Eqs.~\eqref{eq:alphaalpha def}--\eqref{eq:zeta_c def}.
Substituting these into Eqs.~\eqref{eq: P22} and~\eqref{eq: P13} and using Wick's
theorem, we obtain
 \begin{align}
    P^{(22)}_{\delta\delta}(t,{\bm k}) &=
    \frac{2}{(2\pi)^{3}}\int {\rm d}^{3}p\, F_{2}^{2}(t,{\bm p},{\bm k}-{\bm p})P_{\rm L}(p)P_{\rm L}(|{\bm k}-{\bm p}|),
    \label{eq: P^{22}_{dd}}
\\
    P^{(13)}_{\delta\delta}(t,{\bm k}) &=
    \frac{3}{(2\pi)^{3}}P_{\rm L}(k)\int {\rm d}^{3}p\, F_{3}(t,{\bm k},{\bm p},-{\bm p})P_{\rm L}(p).
    \label{eq: P^{13}_{dd}}
\end{align}
Performing the integrals, we arrive at the final form of the one-loop corrections:
\begin{align}
    P^{(22)}_{\delta\delta}(t,{\bm k}) 
    &= \frac{k^2}{(2\pi)^{2}}\int^{\infty}_{0} {\rm d}p\, {\cal P}_{22}(p),
\label{eq: resultant P^{22}_{dd}}
\\
    P^{(13)}_{\delta\delta}(t,{\bm k}) 
    &=\frac{k^2}{(2\pi)^{2}}\int^{\infty}_{0} {\rm d}p\, {\cal P}_{13}(p).
    \label{eq: resultant P^{13}_{dd}}
\end{align}
Here we have defined the kernel functions as
\begin{align}
	{\cal P}_{22}(p)
		&= 
		\frac{P_{\rm L}(p)}{98}\int^{1}_{-1}{\rm d}x\,
			P_{\rm L}\left((k^{2}+p^{2}-2kpx)^{1/2}\right)
			\frac{\left[(7 \kappa -4 \lambda)p +7 \kappa kx -2px^2 (7 \kappa -2 \lambda)\right] ^2}{\left(k^2-2kp x+p^{2}\right)^2}
	,\label{eq: integrand22}\\
	{\cal P}_{13}(p)
		&=\frac{P_{\rm L}(k)P_{\rm L}(p)}{12}
			\Biggl[
				\frac{2}{7}d_{\gamma\gamma}\frac{k^2}{p^2}
				+4\left(\frac{3}{4}{\cal D}-\frac{4}{21}d_{\gamma\gamma}-d_{\alpha\alpha}-d_{\alpha\alpha\ominus}-2d_\zeta\right)
	\notag\\
	&\quad
			+8\left({\cal D}-\frac{1}{28}d_{\gamma\gamma}-d_{\alpha\alpha\ominus}-d_\zeta\right)\frac{p^2}{k^2}
			+3{\cal D}\,\frac{p^4}{k^4}
			+\frac{3}{2}\left( {\cal D}\,\frac{p^2}{k^2}+\frac{2}{21}d_{\gamma\gamma}\right)\frac{(k^2-p^2)^3}{k^3p^3}\ln\left(\frac{k+p}{|k-p|}\right)
			\Biggr]
	,
\label{eq: integrand13}
\end{align}
where $x$ denotes the directional cosine between ${\bm k}$ and ${\bm p}$ defined as $x={\bm k}\cdot{\bm p}/kp$,
and we have introduced
\begin{align}
	{\cal D}:=d_{\alpha\alpha}-\frac{4}{7}d_{\alpha\gamma}-\frac{2}{21}d_{\gamma\gamma}-d_{\alpha\alpha\ominus}-d_{\alpha\gamma\ominus}
	.
\end{align}
Given a concrete
model of modified gravity and a linear power spectrum,
it is now straightforward to
calculate the one-loop matter power spectrum using
Eqs.~(\ref{eq: resultant P^{22}_{dd}}) and~(\ref{eq: resultant P^{13}_{dd}}).

\subsection{Asymptotic behaviors of the loop integrals}\label{sec: loop}

In order to study the one-loop contributions to the matter power spectrum in the context of DHOST theories, we would like to examine
their asymptotic behavior of the short and long wavelength limits in the loop integrals
as done in Ref.~\cite{Makino:1991rp}.
To do this, let us divide the one-loop contributions into that from the momentum integration for $p\gg k$ (UV region) 
and that from the integration for $p\ll k$ (IR region), 
for fixed $k$. 
It was shown in Ref.~\cite{Makino:1991rp} that, when assuming GR and the standard linear matter power 
spectrum,\footnote{
Assuming the scale-invariant primordial curvature perturbations, 
the standard scale dependence of the linear power spectrum is roughly given by
\begin{align}
    P_{\rm L}(k) \propto kT^{2}(k) \propto 
    \begin{cases}
    k\hspace{0.65cm}(k\ll k_{\rm eq}),\\
    k^{-3}\quad(k\gg k_{\rm eq}),
    \end{cases}
\end{align}
where $T(k)$ is the transfer function and $k_{\rm eq}$ is the wave-number at the matter-radiation equality time.}
the leading terms from ${\cal P}_{22}$ and ${\cal P}_{13}$ in the IR limit are exactly canceled out 
and the loop integrals in both the IR and UV regions are convergent.
In this section, we extend their analysis to DHOST theories, and in particular, we investigate the asymptotic behavior 
of the matter power spectrum and the condition for their convergence.

Let us first consider the long-wavelength contribution in the IR limit, namely $p/k\to 0$. 
In the naive $p\to 0$ limit of Eq.~\eqref{eq: integrand22}, we have
\begin{align}
	{\cal P}_{22}\to \frac{1}{3}\kappa^2P_{\rm L}(k)P_{\rm L}(p).
    \label{eq: P22 IR}
\end{align}
However, as pointed out in Ref.~\cite{Makino:1991rp}, since the second-order kernel $F_2(t,{\bm p},{\bm k}-{\bm p})$ is 
symmetric between ${\bm p}$ and ${\bm k}-{\bm p}$,
we also have to take into account of the $|{\bm k} -{\bm p}|\to0$ limit so that the integrand in the appropriate limit is twice larger than Eq.~(\ref{eq: P22 IR}).
Hence, the appropriate IR limit of \eqref{eq: integrand22} is given by
\begin{align}
	{\cal P}_{22}\to \frac{2}{3}\kappa^2P_{\rm L}(k)P_{\rm L}(p).
	\label{eq:P22 IR correct}
\end{align}
On the other hand, the same limit of Eq.~\eqref{eq: integrand13} yields
\begin{align}
	{\cal P}_{13}\to -\frac{1}{3}\left(d_{\alpha\alpha} +d_{\alpha\alpha\ominus} +2d_{\zeta}\right) P_{\rm L}(k)P_{\rm L}(p)
	.\label{eq:P13 IR}
\end{align}
We find from Eqs.~\eqref{eq:P22 IR correct} and \eqref{eq:P13 IR} 
that 
in the IR limit the sum of the kernel functions in the one-loop correction,
${\cal P}_{22}+2{\cal P}_{13}$,
is cancelled out 
within the Horndeski family of theories, $\kappa =d_{\alpha\alpha}=1$ and $d_{\alpha\alpha\ominus}=d_{\zeta}=0$, 
and the convergence condition is the same as those in GR.
However, once we consider DHOST theories, this cancellation does not occur, and the convergence condition of the loop integrals seems to become more stringent than that of standard one in GR. 
This phenomenon has been already suggested in Refs.~\cite{Crisostomi:2019vhj,Lewandowski:2019txi}, but we derived the explicit form of the asymptotic behavior of ${\cal P}_{22} + 2{\cal P}_{13}$ in terms of the functions characterizing DHOST theories.
We anticipate 
that the more stringent convergence condition of the loop integrals in the IR region could originate from strong correlations between short and long modes
in such theories. In order to support  
this, in Appendix \ref{sec: trispectrum}, we revisit the matter trispectrum
in DHOST theories and look at the several limiting cases.

Let us move on the short-wavelength contribution in the UV limit, namely $p/k\to\infty$.
Hereafter, we assume that the linear power spectrum $P_{\rm L}(p)$ in the UV regions behaves asymptotically in
proportion to $p^n$.
In the $p/k \to\infty$ limit, Eq.~\eqref{eq: integrand22} reduces to 
\begin{align}
	{\cal P}_{22}\to\frac{343\kappa^2-336\kappa\lambda+128\lambda^2}{735}\frac{k^2}{p^2}\Bigl[ P_{\rm L}(p)\Bigr]^2
	\propto p^{2(n-1)}
	.
\end{align}
This expression can be rewritten as
\begin{align}
	\frac{P^{(22)}_{\delta\delta}(k)}{P_{\rm L}(k)}
		\to\frac{343\kappa^2-336\kappa\lambda+128\lambda^2}{735(2\pi)^2}\frac{k^4}{P_{\rm L}(k)}\int_{p\gtrsim k} {\rm d}p\, p^{-2}\Bigl[ P_{\rm L}(p)\Bigr]^2
	.
\end{align}
Thus, this term is convergent for $n\leq 1/2$, which is the same as the convergence condition in GR.
We then investigate the same limit of Eq.~\eqref{eq: integrand13},
\begin{align}
	{\cal P}_{13}\to -\frac{2}{3}\left( d_{\alpha\alpha\ominus}+d_\zeta\right)\frac{p^2}{k^2}P_{\rm L}(k)P_{\rm L}(p)
	\propto p^{n+2}.
    \label{eq:13expansion}
\end{align}
Hence, we have
\begin{align}
	\frac{P^{(13)}_{\delta\delta}(k)}{P_{\rm L}(k)}
		\to -\frac{2\left( d_{\alpha\alpha\ominus}+d_\zeta\right)}{3(2\pi)^2}\int_{p\gtrsim k}{\rm d}p\, p^2P_{\rm L}(p)
	.
\end{align}
which immediately leads to that the
integration with the UV regions
in $P^{(13)}_{\delta\delta}$ is separately convergent only for $n\leq -3$. 
We then find that its leading dependence on $p$ is stronger and
the condition of its convergence becomes more stringent than that in GR.
On the other hand, in the case of the Horndeski theory, the coefficient of the leading term vanishes and 
the next-to-leading term is given by
\begin{align}
	{\cal P}_{13}^{\rm Horn}\to\frac{147-144\lambda -64d_{\gamma\gamma}}{315}P_{\rm L}(k)P_{\rm L}(p)\propto p^n
	,
\end{align}
implying that the convergence condition
reduces to that in GR, $n\leq -1$.
Therefore, we conclude that in DHOST theories beyond Horndeski the linear power spectrum
should be required to be redder than that in the case of the Horndeski theory and GR for the convergence of the one-loop correction.
An important observation is that the standard linear power spectrum which behaves as $P_{\rm L}(k) \propto k^{-3}$
for short wavelengths
is on the edge of the convergence in DHOST theories. 
Note that
the coefficient of the leading term, $d_{\alpha\alpha \ominus}+ d_\zeta$, does not vanish even in the viable DHOST theory evading gravitational wave constraints~\cite{Creminelli:2018xsv,Creminelli:2019kjy}.

The more stringent convergence conditions of the loop integrals could be interpreted from the point of view of quantum field theory.
As usually discussed in quantum field theory, symmetry protects loop corrections of correlation functions. 
As reported in Ref.~\cite{Crisostomi:2019vhj,Lewandowski:2019txi}, Horndeski theories have the accidental symmetry which related to the Friedmann-Lema\^{i}tre-Robertson-Walker symmetry and shift symmetry in terms of fields (see~\cite{Crisostomi:2019vhj,Lewandowski:2019txi} as the detailed discussion) while operators in DHOST theories beyond Horndeski violate that.
So, this violation may be related to the more stringent
convergence conditions of the loop integrals in DHOST theories beyond Horndeski.
Or, moving to the Einstein frame, the coupling between matter and the scalar degree of freedom could be large, so that the prediction based on perturbation theory might not be reliable.

Before closing the section, let us suggest some possibilities to resolve this UV 
sensitive behavior of the one-loop matter power spectrum
in DHOST theories.
The first possibility is,
as we have already discussed, 
to consider the linear power spectrum
with the power-law index being $n \leq -3$
for short wavelengths.
The second is to introduce 
the cut-off scale in the matter power spectrum,
which depends on the nature
of dark matter~\cite{Loeb:2005pm}.
One may also have another, rather different, possibility
that one eliminates the UV terms at the level of the integrand, namely,
one imposes the additional condition $d_{\alpha\alpha\ominus}+d_\zeta =0$, which can be used to add the constraint on the combination of the parameters,
on the basis of the assumption
that this UV divergent behavior would be spurious and must vanish.

\section{Conclusions}\label{sec: conclusion}

In this paper, we have studied the third-order solution of the matter density perturbations
and the one-loop matter power spectrum 
in the context of the degenerate higher-order scalar-tensor (DHOST) theories.
We have solved the field equations for the gravitational potentials and scalar field perturbation
order by order under the quasi-static approximation,
and obtained the formal solutions
at all order.
We then explicitly presented the second- and third-order 
non-linear terms appearing in the evolution equation for the density perturbation.
The second- and third-order solutions can be characterized, respectively, by two and seven
functions
describing the non-linear mode-couplings.
In particular, we found that at third order there appear
three new shape functions in the momentum space [Eqs.~\eqref{eq:alphaalpha- def}, \eqref{eq:alphagamma- def},
and~\eqref{eq:zeta_c def}]
in DHOST theories beyond Horndeski,
which could yield the unique signature of
this new class of scalar-tensor theories. 

Furthermore, by using the resultant second- and third-order solutions of the matter density perturbations,
we calculated the one-loop matter power spectrum and
investigated their asymptotic behavior in the short and long wavelength limits in the loop integrals.
Although as far as the Horndeski theory is concerned the asymptotic behavior both in the infrared (IR) and ultraviolet (UV) limits is basically the same as that in general relativity,
we have shown that
in DHOST theories
the behavior of the loop integrals can be drastically changed.
At the IR limit, the leading terms in $P^{(22)}_{\delta\delta}$ and $P^{(13)}_{\delta\delta}$ do not cancel and the condition for the IR convergence
is thus more stringent than the standard one in general relativity.
Even though this feature has been already discussed in Refs~\cite{Crisostomi:2019vhj,Lewandowski:2019txi},
we derive the complete expressions for the leading terms in terms of the functions
characterizing the theories
and it can make the origin
of this distinctive IR behavior
in DHOST theories clearer.
As discussed in Appendix \ref{sec: trispectrum},
we anticipate that 
the more stringent convergence condition of the loop integrals in the IR limit could originate 
from strong correlations between short and long modes in such theories.
For the UV limit, 
we have shown that
the loop integral related to the third-order solution in DHOST theories
has logarithmic divergence
in the case of
the standard linear power spectrum.
Hence, we conclude that the one-loop contributions to the matter power spectrum would be sensitive to the short-wavelength behavior of the linear power spectrum
as long 
as gravity is described by DHOST theories
beyond Horndeski.

\acknowledgements

This work was supported in part by
JSPS KAKENHI Grant Nos.
JP19H01895 (S.H.),
JP20H04745 (T.K.), JP20K03936 (T.K.),
JP17K14304 (D.Y.), 19H01891 (D.Y.), JP20K03968 (S.Y.), 
and JP20H01932 (S.Y.).

\appendix

\section{Definition of the coefficients in the equations of motion}
\label{sec:Definition of the coefficients in the equations of motion}

In this section, we summarize the definition of the effective field theory parameters and
the coefficients in the equations of motion.
In addition to the parameters that appear in the class Ia degeneracy conditions~(\ref{eq: degeneracy1}) and~(\ref{eq: degeneracy2}), one can characterize cosmological perturbations in DHOST theories by introducing $\alpha_{\rm B}$, $\alpha_{\rm M}$, and $\alpha_{\rm V}$ defined by
\begin{align}
    M^{2}H\alpha_{\rm M} &= 
    \frac{\D}{\D t}M^{2},
\\
    M^2H\alpha_{\rm B} &=  
    M^{2}H\alpha_{\rm V} -3M^2H\beta_{1} 
    +\dot\phi\left(-XG_{3X} +G_{4\phi } +2X G_{4\phi X}\right) 
\notag\\
    &\quad +\dot\phi\ddot\phi [2 X\left(G_{4XX} -a_{2X} +Xa_{3X} -a_{4} +2X a_{5}\right) 
    +3(G_{4X} -a_{2} +X a_{3})],
\\
    M^{2}\alpha_{\rm V} &= 4X(G_{4X} -2a_{2} -2Xa_{2X}).
\end{align}
These parameters appear within Horndeski theories.
Note that we have yet another parameter which is often denoted as $\alpha_{\rm K}$, but it does not appear in the equations under the quasi-static region
({\em i.e.}, on sub-horizon scales).

The explicit expressions of the coefficients in Eqs.~(\ref{eq: EoM Psi F})--(\ref{eq: EoM Q F}) are given by
\begin{align}
    c_1 &= 
    -4\left[\alpha_{\rm B} -\alpha_{\rm H} +\frac{\beta_3}{2}(1+\alpha_{\rm M}) +\frac{\dot\beta_3}{2H}\right],
\\
    c_2 &= 
    4\left[\alpha_{\rm H}(1+\alpha_{\rm M}) +\alpha_{\rm M} -\alpha_{\rm T} +\frac{\dot\alpha_{\rm H}}{H}\right],
\\
    c_3 &= 
    -2\Bigg\{ \left(1 +\alpha_{\rm M} +\frac{\dot H}{H^2}\right)\left(\alpha_{\rm B} -\alpha_{\rm H}\right)
        +\frac{\dot\alpha_{\rm B} -\dot\alpha_{\rm H}}{H} +\frac{3\Omega_{\rm m}}{2} +\frac{\dot H}{H^2} +\alpha_{\rm T} -\alpha_{\rm M}
\\
    &\quad +\left[-2\frac{\dot H}{H^2}\beta_{1}  +\frac{\beta_3}{4}(1 +\alpha_{\rm M}) +\frac{\dot\beta_3}{2H}\right]\left(1 +\alpha_{\rm M} -\frac{\dot H}{H^2}\right) -2\frac{\dot H}{H^2}\frac{\dot\beta_1}{H}
        +\left(\frac{\dot H}{H^2}\right)^2\frac{\beta_3}{2} +\frac{\dot\alpha_{\rm M}}{H}\frac{\beta_3}{4} +\frac{\ddot\beta_3}{4H^2}  \Bigg\},
\\
    b_{1} &= \frac{c_1}{4}+\frac{1}{2}(1+\alpha_{\rm M})(2\beta_1+\beta_3)+\frac{1}{2}
\frac{\D}{\D t}\left(\frac{2\beta_1+\beta_3}{H}\right),
\\
    b_2 &=
    -\frac{c_2}{4}+(1+\alpha_{\rm M})\alpha_{\rm H}+
\left(\frac{\alpha_{\rm H}}{H}\right)^{\dd},
\\
    b_3 &= 
    2c_3+\left[
    \left(1+\alpha_{\rm M}-\frac{\dot H}{H^2}\right)(1+\alpha_{\rm M})+\frac{\dot \alpha_{\rm M}}{H}\right](4\beta_1+\beta_3)
    +2(1+\alpha_{\rm M})\left(\frac{4\beta_1+\beta_3}{H}\right)^{\dd}
    +\left(\frac{2\beta_1+\beta_3}{H^2}\right)^{\dd\dd},\label{eq:b3}
\\
    b_4 &= 
    2\Biggl[\left(1+\alpha_{\rm M}-\frac{\dot H}{H^2}\right)(4\beta_1+\beta_3)
    +\left(\frac{4\beta_1+\beta_3}{H}\right)^{\dd}\Biggr],
\\
    d_{1} &=
    -\left[\alpha_{\rm V} +3(\alpha_{\rm H} -\alpha_{\rm T}) -4\alpha_{\rm B}
        +\alpha_{\rm M}(2 -\alpha_{\rm V} +\alpha_{\rm H} +8\beta_{1})
        +2(4\beta_{1} +\beta_{3})\frac{\dot H}{H^{2}}
        -\frac{\dot \alpha_{\rm V} -\dot\alpha_{\rm H} -8\dot\beta_{1}}{H}\right],
\\
    d_{2} &=
    \frac{1}{2}(\alpha_{\rm V} -\alpha_{\rm H}-4\beta_{1}),
\end{align}
where $\Omega_{\rm m}$ was defined in Eq.~\eqref{eq:Omega_m def}.

\section{Coefficients of first-, second- and third-order solutions}

In this section, we summarize the coefficients of first-, second- and third-order solutions.

\subsection{Homogeneous solutions}\label{sec: coeff homogeneous}

The components of the matrices $\mathsf{M}$ and $\mathsf{N}$ in Eq.~(\ref{eq: psiphi}) are read off from Eqs.~(\ref{eq: EoM Psi F}),~(\ref{eq: EoM Phi F}) as
\begin{align}
    \mathsf{M} &
    =\left(\mathsf{M}\right)_{ab}
    =\left(\begin{array}{cc}
    \mathsf{M}_{\Psi\Psi} & \mathsf{M}_{\Psi\Phi} \\
    \mathsf{M}_{\Phi\Psi} & \mathsf{M}_{\Phi\Phi} \\
    \end{array}\right)
    = \left(\begin{array}{cc}
    1+\alpha_{\rm T} & -(1+\alpha_{\rm H}) \\
    1+\alpha_{\rm H} &-\beta_{3}/2 \\
    \end{array}\right),
\\
    \mathsf{N} &
        =\left(\begin{array}{cc}
    \mathsf{N}_{\Psi Q} & \mathsf{N}_{\Psi \dot Q} \\
    \mathsf{N}_{\Phi Q} & \mathsf{N}_{\Phi \dot Q} \\
    \end{array}\right)
    = \left(\begin{array}{cc}
    -b_{2} & -\alpha_{\rm H} \\
    -b_{1} & -(2\beta_{1} +\beta_{3})/2 \\
    \end{array}\right),
\end{align}
with $a,b$ stands for $\Psi$ and $\Phi$.
The coefficients in \eqref{eq: Psi F_{n}} and \eqref{eq: Phi F_{n}} can be written in terms of
above quantities and the coefficient of Eq.~\eqref{eq: Q F_{n}} as
\begin{align}
	\mu_{a} &= (\mathsf{M}^{-1}\mathsf{N})_{a\dot Q}\,\nu_{Q},
	\\
	\nu_{a} &= (\mathsf{M}^{-1}\mathsf{N})_{aQ}\,\nu_{Q}  
					+(\mathsf{M}^{-1}\mathsf{N})_{a\dot Q} \left[\kappa_{Q} +\frac{(a^{2}H\nu_{Q})^{\dd}}{a^{2}H^{2}}\right],
	\\
	\kappa_{a} &= \frac{3}{2}\Omega_{\rm m}(\mathsf{M}^{-1})_{a\Phi}  +(\mathsf{M}^{-1}\mathsf{N})_{aQ}\,\kappa_{Q}  
		+(\mathsf{M}^{-1}\mathsf{N})_{a\dot Q} \frac{(a^{2}H^{2}\kappa_{Q})^{\dd}}{a^{2}H^{3}}.
\end{align}
Substituting these back into Eq.~\eqref{eq: EoM Q F}, we obtain the explicit forms of the coefficients in Eq.~(\ref{eq: Q F_{n}}) as
\begin{align}
    \nu_{Q} 
		&=-\frac{3}{2}\frac{\Omega_{\rm m}}{Z}
				\Bigl[
					4\alpha_{\rm H}(\mathsf{M}^{-1})_{\Psi\Phi} -2(2\beta_{1} +\beta_{3})(\mathsf{M}^{-1})_{\Phi\Phi}
				\Bigr],
\\
	\kappa_{Q} 
		&= -\frac{3}{2}\frac{\Omega_{\rm m}}{Z}
				\left[
					c_{2}(\mathsf{M}^{-1})_{\Psi\Phi} +c_{1}(\mathsf{M}^{-1})_{\Phi\Phi}  
					+\frac{4aM^2\alpha_{\rm H}}{H}\left[\frac{1}{aM^2}(\mathsf{M}^{-1})_{\Psi\Phi}\right]^{\dd}  
					-\frac{2aM^2(2\beta_{1}+\beta_{3})}{H}\left[\frac{1}{aM^2}(\mathsf{M}^{-1})_{\Phi\Phi}\right]^{\dd}
				\right],
\end{align}
where 
\begin{align}
	Z &=b_{3} +c_{2}(\mathsf{M}^{-1}\mathsf{N})_{\Psi Q} 
			+c_{1}(\mathsf{M}^{-1}\mathsf{N})_{\Phi Q} 
			+\frac{4\alpha_{\rm H}}{H}\left[(\mathsf{M}^{-1}\mathsf{N})_{\Psi Q}\right]^{\dd} 
			-\frac{2(2\beta_{1}+\beta_{3})}{H}\left[(\mathsf{M}^{-1}\mathsf{N})_{\Phi Q}\right]^{\dd}
	.\label{eq:Z def}
\end{align}

\subsection{General expression of coefficients of higher-order solutions}
\label{sec:Coefficients of higher-order solutions}

We show that the $n$-th order coefficient with the shape $\Pi$ in Eqs.~\eqref{eq: Psi F_{n}} and \eqref{eq: Phi F_{n}}
is generally written in terms of the $n$-th order coefficient of $Q^{(n)}$ and the matrix components of
$\mathsf{M}$, $\mathsf{N}$, and $\mathsf{O}^{(n)}$ as
\begin{align}
	\tau^{(n)}_{a,\Pi} 
		&=(\mathsf{M}^{-1}\mathsf{O}^{(n)})_{a\Pi}+(\mathsf{M}^{-1}\mathsf{N})_{aQ}\tau^{(n)}_{Q,\Pi} 
			+(\mathsf{M}^{-1}\mathsf{N})_{a\dot Q}\frac{(a^{2}H^{2}\tau^{(n)}_{Q,\Pi})^{\dd}}{a^{2}H^{3}}
	.
\end{align}
Substituting the $n$-th order solutions of $\Psi$, $\Phi$ and Eq.~\eqref{eq: Q F_{n}} into Eq.~\eqref{eq: EoM Q F},
we then obtain the form of $\tau_{Q,\Pi}^{(n)}$ as
\begin{align}
	\tau^{(n)}_{Q,\Pi} 
		&=\frac{1}{Z}\biggl[
			\mathsf{O}^{(n)}_{Q,\Pi} 
			-c_{2}(\mathsf{M}^{-1}\mathsf{O}^{(n)})_{\Psi\Pi}
			-c_{1}(\mathsf{M}^{-1}\mathsf{O}^{(n)})_{\Psi\Pi}
	\notag\\
	&\quad  
			-\frac{4\alpha_{\rm H}}{a^{2}H^{3}}\left[ a^{2}H^{2}(\mathsf{M}^{-1}\mathsf{O}^{(n)})_{\Psi\Pi}\right]^{\dd}
			+\frac{2(2\beta_1+\beta_3)}{a^{2}H^{3}}\left[ a^{2}H^{2}(\mathsf{M}^{-1}\mathsf{O}^{(n)})_{\Phi\Pi}\right]^{\dd}
		\biggr]
	,\label{eq: tauQ2}
\end{align}
where $Z$ was defined in Eq.~\eqref{eq:Z def}.
Here, the coefficient $\mathsf{O}^{(n)}_{Q,\Pi}$
EoM of $Q$ and the coefficient of $W_\Pi^{(n)}$.
Therefore, once the lower-order solutions and the $n$-th order matrix components of $\mathsf{O}^{(n)}$ are given, 
we can straightforwardly derive the $n$-th order solution of $\Psi$, $\Phi$, and $Q$.

\subsection{Second-order solutions}\label{sec: coeff second}

To derive the second-order coefficients in Eqs.~\eqref{eq: Q F_{n}}--\eqref{eq: Phi F_{n}}, we need to 
write down the reduced first order solution. 
When substituting Eqs.~\eqref{eq: first-order eq.} and \eqref{eq:f def} into the first-order solution 
of Eqs.~\eqref{eq: Q F_{n}}--\eqref{eq: Phi F_{n}},
$\Psi^{(1)}$,
$\Phi^{(1)}$, and $Q^{(1)}$ can be rewritten as
\begin{align}
	&\Psi^{(1)}(t,{\bm p})=-\frac{a^2(t)H^2(t)}{p^2}K_\Psi (t)D_+(t)\delta_{\rm L}({\bm p})
	\,,\label{eq:reduced Psi^1}\\
	&\Phi^{(1)}(t,{\bm p})=-\frac{a^2(t)H^2(t)}{p^2}K_\Phi (t)D_+(t)\delta_{\rm L}({\bm p})
	,\\
	&Q^{(1)}(t,{\bm p})=-\frac{a^2(t)H^2(t)}{p^2}K_Q (t)D_+(t)\delta_{\rm L}({\bm p})
	\,,\label{eq:reduced Q^1}\\
	&\dot Q^{(1)}(t,{\bm p})=-\frac{a^2(t)H^3(t)}{p^2}K_{\dot Q} (t)D_+(t)\delta_{\rm L}({\bm p})
	,\label{eq:reduced dot Q^1}
\end{align}
At the second-order, the relevant shape functions to describe the solutions
are shown to be $\alpha_s ({\bm k}_1,{\bm k}_2)$  and $\gamma ({\bm k}_1,{\bm k}_2)$,
which are defined in Eqs.~\eqref{eq: alphas} and \eqref{eq: gamma}.
Since the non-linear interaction in Eqs.~\eqref{eq: EoM Psi F} and \eqref{eq: EoM Phi F} are determined by $Q$,
the matrix components of $\mathsf{O}$ in Eq.~\eqref{eq: psiphi} at the second-order, that is $\mathsf{O}^{(2)}$, 
can be written in terms of the first order solution of $Q$. We then have
\begin{align}
	\mathsf{O}^{(2)} 
		=\left(
			\begin{array}{cc}
			\mathsf{O}^{(2)}_{\Psi,\alpha}& \mathsf{O}^{(2)}_{\Psi,\gamma} \\
			\mathsf{O}^{(2)}_{\Phi,\alpha} &  \mathsf{O}^{(2)}_{\Phi,\gamma} \\
			\end{array}
			\right)
		=\frac{1}{4}D_+^2K_Q^2
			\left(
			\begin{array}{cc}
			4\alpha_{\rm H} & \alpha_{\rm T} -4\alpha_{\rm H} \\
			2(2\beta_{1}+\beta_{3}) &  -2(d_{2} +2\beta_{1}+\beta_{3}) \\
			\end{array}
			\right)
	.
\end{align}
Moreover, with the use of $K_\Psi$, $K_\Phi$, and $K_Q$, and the shape functions, the coefficient in Eq.~\eqref{eq: tauQ2}
is given by
\begin{align}
	\mathsf{O}^{(2)}_{Q,\alpha} 
		&= D_+^2K_Q\left\{
			4\alpha_{\rm H}K_{\Psi} 
			-2(2\beta_{1}+\beta_{3})K_{\Phi}
			+b_4K_{Q}
			+6(4\beta_1+\beta_3)K_{\dot Q}
			\right\}
	,\\
	\mathsf{O}^{(2)}_{Q,\gamma} 
		&= -D_+^2K_Q\left\{
			2\alpha_{\rm T}K_{\Psi}
			+4d_{2}K_{\Phi}
			+\left(d_1+b_4\right) K_{Q}
			+4(4\beta_{1}+\beta_{3})K_{\dot Q}
			\right\}
	.
\end{align}

\subsection{Third-order solutions}\label{sec: coeff third}

Following the same step as the previous subsection, 
to derive the third-order solutions, it is useful to introduce the reduced second-order solutions.
Substituting the second-order solution Eq.~\eqref{eq: delta_{2}} into Eqs.~\eqref{eq: Q F_{n}}--\eqref{eq: Phi F_{n}},
we obtain
\begin{align}
	\Psi^{(2)}(t,{\bm p})
		=&-\frac{a^2(t)H^2(t)}{p^2}\biggl[\widetilde\tau_{\Psi ,\alpha}(t)W_\alpha ({\bm p})+\widetilde\tau_{\Psi ,\gamma}(t)W_\gamma ({\bm p})\biggr]
	,\label{eq:reduced Psi^2}\\
	\Phi^{(2)}(t,{\bm p})
		=&-\frac{a^2(t)H^2(t)}{p^2}\biggl[\widetilde\tau_{\Phi ,\alpha}(t)W_\alpha ({\bm p})+\widetilde\tau_{\Phi ,\gamma}(t)W_\gamma ({\bm p})\biggr]
	,\\
	Q^{(2)}(t,{\bm p})
		=&-\frac{a^2(t)H^2(t)}{p^2}\biggl[\widetilde\tau_{Q ,\alpha}(t)W_\alpha ({\bm p})+\widetilde\tau_{Q ,\gamma}(t)W_\gamma ({\bm p})\biggr]
	,\\
	\dot Q^{(2)}(t,{\bm p})
		=&-\frac{a^2(t)H^3(t)}{p^2}\biggl[\widetilde\tau_{\dot Q ,\alpha}(t)W_\alpha ({\bm p})+\widetilde\tau_{\dot Q ,\gamma}(t)W_\gamma ({\bm p})\biggr]
	,\label{eq:reduced dot Q^2}
\end{align}
where $\widetilde\tau_{*,\Pi}$ can be described by the lower-order solutions and $\tau_{*,\Pi}$ itself.

Let us consider the kernels that describes the non-linear mode-coupling of the third-order solutions.
We first define the kernels that are generated by $\alpha_s$ and $\gamma$ as
\begin{align}
	&\alpha\alpha({\bm k}_{1},{\bm k}_{2},{\bm k}_{3}) 
		=\frac{1}{3}\Bigl[
			\alpha_{s}({\bm k}_{1},{\bm k}_{2}+{\bm k}_{3})\alpha_{s}({\bm k}_{2},{\bm k}_{3}) 
     			+2~{\rm perms.}
		\Bigr]
	,\label{eq:alphaalpha def}\\
	&\alpha\gamma({\bm k}_{1},{\bm k}_{2},{\bm k}_{3}) 
		=\frac{1}{3}\Bigl[
			\alpha_{s}({\bm k}_{1},{\bm k}_{2}+{\bm k}_{3})\gamma({\bm k}_{2},{\bm k}_{3}) 
    			+2~{\rm perms.}
			\Bigr]
	,\\
	&\gamma\alpha ({\bm k}_1,{\bm k}_2,{\bm k}_3)
		=\frac{1}{3}\Bigl[
			\gamma ({\bm k}_1,{\bm k}_2+{\bm k}_3)\alpha_s({\bm k}_2,{\bm k}_3)
			+2~{\rm perms}
		\Bigr]
	,\\
	&\gamma\gamma({\bm k}_{1},{\bm k}_{2},{\bm k}_{3}) 
		=\frac{1}{3}\big[
				\gamma({\bm k}_{1},{\bm k}_{2}+{\bm k}_{3})\gamma({\bm k}_{2},{\bm k}_{3}) 
    				+2~{\rm perms.}
			\big]
	.\label{eq:gammagamma def}
\end{align}
In solving Eqs.~\eqref{eq: EoM Q F}, \eqref{eq: continuous F}, and \eqref{eq: Eular F}, we need
to introduce the kernels that are generated by the asntisymmetric part of $\alpha$ as well as $\alpha_s$ and $\gamma$.
Hence we define
\begin{align}
	&\alpha\alpha_\ominus({\bm k}_{1},{\bm k}_{2},{\bm k}_{3}) 
		= \frac{1}{6}\Big\{
			\bigl[
				\alpha({\bm k}_{1},{\bm k}_{2}+{\bm k}_{3})-\alpha({\bm k}_{2}+{\bm k}_{3},{\bm k}_{1})
			\bigr]
			\alpha_{s}({\bm k}_{2},{\bm k}_{3}) 
			+2~{\rm perms.}
	\Big\}
	,\label{eq:alphaalpha- def}\\
	&\alpha\gamma_\ominus ({\bm k}_{1},{\bm k}_{2},{\bm k}_{3})
		= \frac{1}{6}\Bigl\{
			\bigl[
				\alpha({\bm k}_{1},{\bm k}_{2}+{\bm k}_{3})-\alpha({\bm k}_{2}+{\bm k}_{3},{\bm k}_{1})
			\bigr]
			\gamma({\bm k}_{2},{\bm k}_{3}) 
			+2~{\rm perms.}
			\Bigr\}
	.\label{eq:alphagamma- def}
\end{align}
In addition to these six kernels, we have to take into account the mode-couplings from the three-point self-interaction terms of $Q$ in Eq.~\eqref{eq: EoM Q F},
that is $\xi$ and $\zeta$.
We then define the following cyclic-symmetrized mode-coupling functions as
\begin{align}
    &\xi_{c}({\bm k}_{1},{\bm k}_{2},{\bm k}_{3}) =
    \frac{1}{3}\Bigl\{ \xi({\bm k}_{1},{\bm k}_{2},{\bm k}_{3}) +2~{\rm perms.}\Bigr\}
    ,\\
    &\zeta_{c}({\bm k}_{1},{\bm k}_{2},{\bm k}_{3}) =
    \frac{1}{3} \Bigl\{\zeta({\bm k}_{1},{\bm k}_{2},{\bm k}_{3}) +2~{\rm perms.}\Bigr\}
    .\label{eq:zeta_c def}
\end{align}
In summary, we need to consider the set of the eight kernels, $U_3=\{\alpha\alpha,\alpha\gamma,\gamma\alpha,\gamma\gamma, \alpha\alpha_\ominus,\alpha\gamma_\ominus,\xi_c,\zeta_c\}$. 
With these kernels, the matrix components of $\mathsf{O}^{(3)}$ can be written as
\begin{align}
	&\mathsf{O}^{(3)} 
	=\left(
		\begin{array}{cccccccc}
			\mathsf{O}^{(3)}_{\Psi,\alpha\alpha} & \mathsf{O}^{(3)}_{\Psi,\alpha\gamma} &\mathsf{O}^{(3)}_{\Psi,\gamma\alpha} & \mathsf{O}^{(3)}_{\Psi,\gamma\gamma} & \mathsf{O}^{(3)}_{\Psi,\alpha\alpha\ominus} & \mathsf{O}^{(3)}_{\Psi,\alpha\gamma\ominus} & \mathsf{O}^{(3)}_{\Psi,\xi} & \mathsf{O}^{(3)}_{\Psi,\zeta} \\
    			\mathsf{O}^{(3)}_{\Phi,\alpha\alpha} & \mathsf{O}^{(3)}_{\Phi,\alpha\gamma} &\mathsf{O}^{(3)}_{\Phi,\gamma\alpha} & \mathsf{O}^{(3)}_{\Phi,\gamma\gamma} & \mathsf{O}^{(3)}_{\Phi,\alpha\alpha\ominus} & \mathsf{O}^{(3)}_{\Phi,\alpha\gamma\ominus} & \mathsf{O}^{(3)}_{\Phi,\xi} & \mathsf{O}^{(3)}_{\Phi,\zeta} \\
		\end{array}
	\right)
	\notag\\
	&\quad =D_+K_Q\left(
		\begin{array}{cccccccc}
			2\alpha_{\rm H}\widetilde\tau_{Q,\alpha} & 2\alpha_{\rm H}\widetilde\tau_{Q,\gamma} & (\alpha_{\rm T}-4\alpha_{\rm H})\widetilde\tau_{Q,\alpha}/2 &  (\alpha_{\rm T}-4\alpha_{\rm H})\widetilde\tau_{Q,\gamma}/2 & 0 & 0 & 0 & 0 \\
    			(2\beta_1+\beta_3)\widetilde\tau_{Q,\alpha} & (2\beta_1+\beta_3)\widetilde\tau_{Q,\gamma} & -(d_2+4\beta_1+2\beta_3)\widetilde\tau_{Q,\alpha} & -(d_2+4\beta_1+2\beta_3)\widetilde\tau_{Q,\gamma}& 0 & 0 & 0 & 0 \\
		\end{array}
		\right)
	.
\end{align}
Substituting the reduced lower-order solutions Eqs.~\eqref{eq:reduced Psi^1}--\eqref{eq:reduced dot Q^1} 
and \eqref{eq:reduced Psi^2}--\eqref{eq:reduced dot Q^2} into Eq.~\eqref{eq: EoM Q F}, we can extract
the correspondence between the coefficient $\mathsf{O}^{(3)}_{Q,\Pi}$ and other parameters, which are 
given by
\begin{align}
	&\mathsf{O}^{(3)}_{Q,\alpha\alpha}
		=2D_+\left\{
				4\alpha_{\rm H}K_{(\Psi }
				-2(2\beta_1+\beta_3)K_{(\Phi}
				+b_4K_{(Q}
				+6(4\beta_1+\beta_3) K_{(\dot Q}
			\right\}\widetilde\tau_{Q),\alpha}
	,\\
	&\mathsf{O}^{(3)}_{Q,\alpha\gamma}
		=2D_+\left\{
				4\alpha_{\rm H}K_{(\Psi }
				-2(2\beta_1+\beta_3)K_{(\Phi}
				+b_4K_{(Q}
				+6(4\beta_1+\beta_3) K_{(\dot Q}
			\right\}\widetilde\tau_{Q),\gamma}
	,\\
	&\mathsf{O}^{(3)}_{Q,\gamma\alpha}
		=-2D_+\left\{
				2\alpha_{\rm T}K_{(\Psi }
				+4d_2K_{(\Phi}
				+(d_1+b_4)K_{(Q}
				+4(4\beta_1+\beta_3)K_{(\dot Q}
			\right\}\widetilde\tau_{Q),\alpha}
	,\\
	&\mathsf{O}^{(3)}_{Q,\gamma\gamma}
		=-2D_+\left\{
				2\alpha_{\rm T}K_{(\Psi }
				+4d_2K_{(\Phi}
				+(d_1+b_4)K_{(Q}
				+4(4\beta_1+\beta_3)K_{(\dot Q}
			\right\}\widetilde\tau_{Q),\gamma}
	,
\end{align}
where we have used the round bracket as the symmetrized symbol defined as $K_{(A}\widetilde\tau_{B),\Pi}=(K_A\widetilde\tau_{B,\Pi}+K_B\widetilde\tau_{A,\Pi})/2$.
Introducing the antisymmetric symbol $K_{[A}\widetilde\tau_{B],\Pi}=(K_A\widetilde\tau_{B,\Pi}-K_B\widetilde\tau_{A,\Pi})/2$,
the remaining coefficients can be given by
\begin{align}
	&\mathsf{O}^{(3)}_{Q,\alpha\alpha\ominus}
		=Z\tau_{Q,\alpha\alpha\ominus}
		=4D_+\left\{
				2\alpha_{\rm H}K_{[\Psi }
				-(2\beta_1+\beta_3)K_{[\Phi}
				-(4\beta_1+\beta_3) K_{[\dot Q}
			\right\}\widetilde\tau_{Q],\alpha}
	,\\
	&\mathsf{O}^{(3)}_{Q,\alpha\gamma\ominus}
		=Z\tau_{Q,\alpha\gamma\ominus}
		=4D_+\left\{
				2\alpha_{\rm H}K_{[\Psi }
				-(2\beta_1+\beta_3)K_{[\Phi}
				-(4\beta_1+\beta_3) K_{[\dot Q}
			\right\}\widetilde\tau_{Q],\gamma}
	,\\
	&\mathsf{O}^{(3)}_{Q,\xi}
		=Z\tau_{Q,\xi}^{(3)}
		=D_+^3K_Q^3(2d_2+\alpha_T)
	,\\
	&\mathsf{O}^{(3)}_{Q,\zeta}
		=Z\tau_{Q,\zeta}^{(3)}
		=-2D_+^3K_Q^3(4\beta_1+\beta_3)
	.
\end{align}

\section{Matter trispectrum}\label{sec: trispectrum}

In Sec.~\ref{sec: loop}, we have showed the 
the more stringent convergence condition of the loop integrals in the IR region.
This could be caused by the strong correlations between the short and long modes in  DHOST theories.
To study this, we investigate the matter trispectrum and its several limit.

The trispectrum for $\delta$ is given
in terms of
the four-point correlation function as
\begin{align}
    \langle \delta(t,{\bm k}_{1})\delta(t,{\bm k}_{2})\delta(t,{\bm k}_{3})\delta(t,{\bm k}_{4}) \rangle 
        &= (2\pi)^{3}\delta({\bm k}_{1}+{\bm k}_{2}+{\bm k}_{3}+{\bm k}_{4})T(t,{\bm k}_{1},{\bm k}_{2},{\bm k}_{3},{\bm k}_{4}).
\end{align}
Since the linear density field is assumed to be Gaussian,
the matter trispectrum is given to leading order by
\begin{align}
    T(t,{\bm k}_{1},{\bm k}_{2},{\bm k}_{3},{\bm k}_{4}) &\simeq D^6_+(t) \Bigl[T^{(1122)}(t,{\bm k}_{1},{\bm k}_{2},{\bm k}_{3},{\bm k}_{4})
    +T^{(1113)}(t,{\bm k}_{1},{\bm k}_{2},{\bm k}_{3},{\bm k}_{4})\Bigr],
\end{align}
where
\begin{align}
    T^{(1122)} 
    &= 4P_{\rm L}(k_1)P_{\rm L}(k_2)
    \Big [P_{\rm L}(|{\bm k}_1+{\bm k}_3|)F_2(t,{\bm k}_1,-{\bm k}_1-{\bm k}_3)F_2(t,{\bm k}_2,{\bm k}_1+{\bm k}_3) 
\notag\\
    &\hspace{3cm} +P_{\rm L}(|{\bm k}_1+{\bm k}_4|)F_2(t,{\bm k}_1,-{\bm k}_1-{\bm k}_4)F_2(t,{\bm k}_2,{\bm k}_1+{\bm k}_4)\Big]
    \notag\\
    &\quad +5~{\rm perms.},
    \label{eq: T1122}
\\
    T^{(1113)} 
    &= 6P_{\rm L}(k_1)P_{\rm L}(k_2)P_{\rm L}(k_3)F_3(t,{\bm k}_1,{\bm k}_2,{\bm k}_3) 
    +3~{\rm perms.}.
    \label{eq: T1113}
\end{align}
As we are interested in the interactions between short and long modes,
we take the double soft limit in which two wave vectors are
taken to be much smaller than the other two.
Let us look at the dimensionless reduced trispectrum defined by
 \begin{align}
    Q(t,{\bm k}_{1},{\bm k}_{2},{\bm q}_{1},{\bm q}_{2}) 
    &= \frac{T(t,{\bm k}_{1},{\bm k}_{2},{\bm q}_{1},
    {\bm q}_{2})}{D^6_+(t)[P_{\rm L}(k_1)P_{\rm L}(k_2)P_{\rm L}(q_1)
    +3~{\rm perms.}]}.
\end{align}
In the double soft limit,
$q_1,q_2\ll k_1,k_2$ with
${\bm k}_{1} \approx -{\bm k}_{2}$ and ${\bm q}_{1} \approx -{\bm q}_{2}$,
Eqs.~(\ref{eq: T1122}) and (\ref{eq: T1113}) reduces to
\begin{align}
    T^{(1122)} &\to 8P_{\rm L}(k_{1})P_{\rm L}^{2}(q_{1})\kappa^{2}(t)\alpha_{s}^2({\bm q}_{1},{\bm k}_{1}),
\\
    T^{(1113)} &\to 12P_{\rm L}(k_{1})P_{\rm L}^{2}(q_{1})\Bigl[d_{\alpha\alpha}(t)\alpha\alpha({\bm k}_{1},{\bm q}_{1},{\bm q}_{2}) 
    +d_{\alpha\alpha\ominus}(t)\alpha\alpha_{\ominus}({\bm k}_{1},{\bm q}_{1},{\bm q}_{2})
    +d_{\zeta}(t)\zeta_{c}({\bm k}_{1},{\bm q}_{1},{\bm q}_{2})\Bigr],
\end{align}
and hence the reduced trispectrum reads
\begin{align}
    Q(t,{\bm k}_{1},{\bm k}_{2},{\bm q}_{1},{\bm q}_{2}) 
    \to 
    \frac{P_{\rm L}(q_{1})}{P_{\rm L}(k_{1})}\left(\kappa^{2} -d_{\alpha\alpha} -d_{\alpha\alpha\ominus} -2d_{\zeta}\right)\left(\frac{{\bm q}_{1}\cdot{\bm k}_{1}}{q_{1}^{2}}\right)^2.
    \label{eq: tri DS}
\end{align}
In the Horndeski theory, we have
$\kappa=d_{\alpha\alpha} =1$ and $d_{\alpha\alpha\ominus}=d_{\zeta}=0$,
so that the above would-be leading contribution vanishes.
However, in theories more general than Horndeski,
the above expression does not vanish in general.
We thus see that in the trispectrum there is a non-negligible contribution
in the double soft limit
that appears for the first time in DHOST theories beyond Horndeski.
This result is consistent with the more stringent convergence condition
of the loop integrals in the infrared limit discussed in Sec.~\ref{sec: loop},
and we reproduce the results of Ref.~\cite{Crisostomi:2019vhj,Lewandowski:2019txi}, 
but we derived the explicit form of the matter trispectrum in the double soft limit in terms of the functions characterizing DHOST theories beyond Horndeski.

Before closing this section let us mention some other limits of the trispectrum.
In the crushed limit
(${\bm k}_{1} \simeq{\bm k}_{2} \simeq -{\bm q}_{1} \simeq -{\bm q}_{2}$)
and in the folded limit
(${\bm k}_{1} \simeq -{\bm k}_{2} \simeq {\bm q}_{1} \simeq -{\bm q}_{2}$),
we find no such contributions that appear for the first time
in DHOST theories beyond Horndeski.
The impact of the third-order solution does not appear on the trispectrum in contrast to the one-loop power spectrum.


\end{document}